\documentclass[a4paper,fleqn,usenatbib]{mnras}
\makeatletter
\def\set@curr@file#1{%
  \begingroup
    \escapechar\m@ne
    \xdef\@curr@file{\expandafter\string\csname #1\endcsname}%
  \endgroup
}
\def\quote@name#1{"\quote@@name#1\@gobble""}
\def\quote@@name#1"{#1\quote@@name}
\def\unquote@name#1{\quote@@name#1\@gobble"}
\makeatother
\usepackage{graphicx}

\usepackage{newtxtext,newtxmath}

\usepackage[T1]{fontenc}
\usepackage{ae,aecompl}

\usepackage{multirow}
\usepackage{xcolor}
\definecolor{wenge}{rgb}{0.39, 0.33, 0.32}
\newcommand{\Msun}{\ensuremath{\,{\rm M}_\odot}}                        
\newcommand{\Teff}{\ensuremath{T_{\rm eff}}}                            

\title[TYC 3700-1384-1, V1511 Her and V1179 Her]{Photometric analysis of three totally eclipsing W UMa stars with increasing periods: TYC 3700-1384-1, V1511 Her and V1179 Her}
\author[Eric Broens]{Eric Broens$^{1}$\thanks{E-mail:
eric.broens@skynet.be}\\
$^{1}$Vereniging Voor Sterrenkunde, Belgium}
\begin{document}

\date{Accepted 2020 December 19. Received 2020 December 19; in original form 2020 February 25}

\pagerange{\pageref{firstpage}--\pageref{lastpage}} \pubyear{2019}

\maketitle

\label{firstpage}

\begin{abstract}
The first multi-colour light curve models and period studies for the totally eclipsing W UMa stars TYC 3700-1384-1,
V1511 Her and V1179 Her are presented. All three stars are A-subtype W UMa stars of spectral type F. The light curve
solutions show that TYC 3700-1384-1 has a moderately low mass ratio of $q = 0.182 \pm 0.001$ and a degree of overcontact of $f = 49 \%$.
For V1179 Her a mass ratio $q = 0.153 \pm 0.001$ and a degree of overcontact of $f = 48 \%$ is derived. The solution for V1511 Her is inconclusive, however
the mass ratio is expected to be between $0.13 < q < 0.15$.
The evolutionary status is compared with zero-age main sequence stars taking into account energy transfer
from the primary to the secondary component. The primary component of TYC 3700-1384-1 fits well in the main-sequence, while V1179 Her is more evolved.
The period study reveals for all three stars a continuously increasing period at a rate of ${{\mathrm{d}}P/{\mathrm{d}}t} = 6.1\times10^{-7} d yr^{-1}$, ${{\mathrm{d}}P/{\mathrm{d}}t} = 5.0\times10^{-7} d yr^{-1}$
and ${{\mathrm{d}}P/{\mathrm{d}}t} = 9.6\times10^{-7} d yr^{-1}$ for TYC 3700-1384-1, V1511 Her and V1179 Her respectively. The estimated
mass transfer rates derived from these period changes are  $\dot M=1.6 \times 10^{-7} {\Msun} yr^{-1} $ for TYC 3700-1384-1
and $\dot M= 1.9 \times 10^{-7} {\Msun} yr^{-1} $ for V1179 Her.
\end{abstract}


\section{Introduction}
W UMa-type systems are contact binary stars containing two late-type dwarf stars which both fill their Roche lobes and
share a common envelope. The spectral classes generally range from F to K. They are very common, among the 2.1 million variable
stars that are currently included in the International Variable Star Index\footnote{https://www.aavso.org/vsx/} \citep[VSX;][]{watson06},
\textasciitilde$20 \%$ are catalogued as W UMa Stars (EW) and another \textasciitilde$5 \%$ are catalogued as
contact binaries (EC). Most of the latter group are genuine W UMa stars too. The orbital period of W UMa stars is less than about 1 day,
with typical periods of about 0.4 days. The orbital period distribution shows a sharp cut-off at 0.22 days. Few binaries, with spectral type M,
have been discovered with significantly shorter periods then 0.22 days \citep{nefs12,drake14}. \\
\citet{binnendijk70} divided W UMa stars into an A-subtype and W-subtype based on whether, respectively, the more massive component
or the less massive component is eclipsed during the primary eclipse. 
Despite the temperatures of both stars in W UMa-type systems appear to be nearly equal, they have typically very different
masses. The mass ratio can be as small as about 0.07. It proves to be challenging to develop a sound theoretical model
explaining the structure of W UMa stars. Several models have been developed but none of them provide an adequate explanation
for the observed properties. \citet{webbink03} gives an overview of the main structural models. The origin and evolution of W UMa
stars is also still being debated. The widely held view is that they form from detached cool binaries with initial orbital
periods of a few days. By a combination of evolutionary expansion and angular momentum loss arising from the magnetic stellar
wind a contact binary is formed. Possibly Kozai-Lidov cycles, induced by a third body in a hierarchical triple system, shrink
the orbit of an initially wider binary to orbital periods of a few days \citep{eggleton06,fabrycky07}. 
Model calculations suggest that the stars in W UMa systems eventually will merge. W UMa stars are therefore considered
candidate progenitors for the fast rotating FK Com Stars, Blue Stragglers and Red Novae \citep[e.g.][]{bopp81,tylenda11,stepien15}.
\citet{qian06} and \citet{yang15} suggest that W UMa stars with a low mass ratio and a high fill-out factor are particularly good candidates
to evolve into rapidly rotating single stars. \\
Since W UMa-type binary stars are still not well understood, detailed photometric analyses
and orbital period studies of those systems can provide invaluable information for further theoretical studies.
A physical model that satisfactorily explains the energy and mass transfer between the two components doesn't exist yet.
The presently available software models the radiative properties of the contact envelope for each component separately, based
on the component's polar temperature. This causes significant discontinuities in the neck region for contact binaries with
components having pronounced different temperatures. The resulting system parameters might, consequently, not be reliable in such cases \citep{kochoska18}.
This paper presents a multi-colour light curve model and orbital period study for the totally eclipsing W UMa-type systems TYC 3700-1384-1,
V1511 Her and V1179 Her. The primary and secondary eclipse depths for the stars studied in this paper are nearly equal, indicating a small
temperature difference between both components, hence avoiding the problem with models caused by large discontinuities in the neck region.
For totally eclipsing W UMa stars, the astrophysically important mass ratio can be obtained reliably by light curve modelling \citep{terrell05}.

\section{Observations} \label{section:observations}
Multi-colour-photometric observations have been carried out using a 20-cm Schmidt-Cassegrain telescope located in a privately owned
observatory near Mol, Belgium. The telescope is equipped with a SBIG ST7-XMEI CCD camera with photometric filters. A focal reducer
provides a field of view of 17 arcmin by 11 arcmin and a plate scale of 1.32 arcsec pixel$^{-1}$. The small field of view limits the number of
available and suitable comparison stars of similar brightness and colour as the targets. The operating temperature of the CCD is kept constant
at a temperature of about 30\degr C below the ambient temperature, which is between $-$35\degr C and $-$10\degr C depending on the season. The stars
were kept on approximately the same pixels during an observing session by means of the CCD camera's additional auto-guider chip.
Since all target stars are of similar apparent brightness, the integration time for all observations was $90 s$. The SNR of the
comparison stars and target stars is at all times higher than 200. \\
The journal of observations is presented in Table \ref{tabLog}.
The initial observations have been obtained with the aim to derive the stellar parameters by modelling the light curves.
The complete orbital phase is covered more than once. Later on additional observations have been acquired with the aim to obtain more
times of minimum light in order to study the orbital period and to complement and verify the times of minimum light calculated from
publicly available data from sky surveys.
The images were processed with dark removal and flat-field
corrections using the {\sc imred} packages in {\sc iraf}\footnote{{\sc iraf}
is distributed by the National Optical Astronomy Observatories, which are operated by the Association of Universities for Research in
Astronomy, Inc., under cooperative agreement with the National Science Foundation.}. \\
Differential photometry was performed with the {\sc digiphot/apphot} package in {\sc iraf}. The
magnitude and colours of the target and check star, transformed to the standard Johnson-Cousins system, is the average of the values
obtained for each comparison star individually. The standard deviation is on average $\sim 0.01$ magnitude for all photometric measurements
of the targets. The magnitudes and colours for the comparison stars are taken from the AAVSO Photometric All Sky Survey DR9 \citep[APASS;][]{henden14}.
Since APASS does not use Cousins $R_{c}$ and $I_{c}$ filters, the APASS Sloan magnitudes have been transformed to $V-R_{c}$ and $R_{c}-I_{c}$ colours by applying the equations from \citet{jester05}.
Table \ref{tab:tabCoo} provides the magnitudes and colours of the used comparison and check stars for each of the observed stars. For the studied stars
the measured $V$ magnitude and $V-R{c}$ and $V-I{c}$ colour indices at first quadrature phase are listed.
The field around TYC 3700-1384-1 was visited on 10 nights by APASS, resulting in \textasciitilde 40 measurements for each of the comparison stars.
The reported uncertainties for the selected comparison stars are $\sigma_{V} = 0.04$, $\sigma_{r'} = 0.1$ and $\sigma_{i'} = 0.1$. The selected comparison stars in the field of V1179 Her were
measured between 26 and 40 times in total on 5 to 7 nights with uncertainties of $\sigma_{V} = 0.04$, $\sigma_{i'} = 0.15$ and $\sigma_{r'} = 0.04$.
The field of V1511 Her was only observed on 2 nights by APASS, totalling 10 measurements for each comparison star. The listed errors are much smaller, $\sigma_{V} \sol 0.02$, $\sigma_{i'} \approx 0.04$ and
$\sigma_{r'} \approx 0.04$.
The large uncertainties on the Sloan magnitudes for the fields of TYC 3700-1384-1 and V1179 Her suggest that these comparison stars are unusable for accurate photometry.
Fortunately the stars discussed in this paper were among the 606 W UMa stars for which \citet{terrell12} measured colour indices. Their colour indices for TYC 3700-1384-1, $V-R_{c} = 0.46\pm 0.01$, 
$V-I_{c} = 0.91 \pm 0.02$, and for V1179 Her, $V-R_{c} = 0.35\pm 0.01$, $V-I_{c} = 0.69 \pm 0.01$ are in good agreement with the ones presented in this paper, $V-R_{c} = 0.46\pm 0.01$, 
$V-I_{c} = 0.90 \pm 0.02$ and $V-R_{c} = 0.38\pm 0.02$, $V-I_{c} = 0.72 \pm 0.04$ respectively.
For V1511 Her however, the $V-R_{c} = 0.35 \pm 0.01$ and $V-I_{c} = 0.67 \pm 0.01$ colours reported by \citeauthor{terrell12} are significantly redder, although their $B-V = 0.49 \pm 0.01$ colour
index suggest a higher ${\Teff}$ than their $V-I_{c}$ and $V-R_{c}$ indices do. The colour indices from our observations are $V-R_{c} = 0.27 \pm 0.02$ and $V-R_{c} = 0.51 \pm 0.03$.
Since the field of V1511 Her was only observed on two nights by APASS, the magnitudes might be
less accurate than the error estimates indicate. The APASS magnitudes appear to be accurate when they are averaged over a sufficient number of nights.
The comparison stars in the field of TYC 3700-1384-1, with the exception of comparison star 1, are significantly redder than the variable. With the low galactic latitude $b \approx -6 \degr$, 
these comparison stars are not necessarily intrinsically red making them susceptible to variability. All comparison stars used in this study remained constant within the measurement errors during the course of the observations.
As will be shown in section \ref{subsection:TYC 3700-1384-1}, the reddening for TYC 3700-1384-1 is estimated to be $E(B-V) = 0.273$.
Even though the colour response of the used equipment is close to the Johnson-Cousins standard system, the significant colour differences between variable and comparison star can introduce
an additional systematic error when the colour terms of the transformation are not sufficiently accurate. Table \ref{tab:compmatrix} provides a matrix with the averaged $V$ magnitude
and the $V-R_{c}$ and $V-I_{c}$ colour indices for the selected comparison stars in this field as measured from the other ones during one night. This table demonstrates that the
transformations are accurate.

\begin{table*}
\caption{Observation log. \label{tabLog}}
\begin{tabular}{ccccc|ccccc|ccccc}
\hline
\multicolumn{2}{|c|}{TYC 3700-1384-1} & \multicolumn{3}{|c|}{Number of data points} & \multicolumn{2}{|c|}{V1511 Her}  & \multicolumn{3}{|c|}{Number of data points} & \multicolumn{2}{|c|}{V1179 Her}  & \multicolumn{3}{|c|}{Number of data points} \\
date           & hours      &     $V$   & $R_{c}$   & $I_{c}$  &    date           & hours     &   $V$    & $R_{c}$  & $I_{c}$  &    date           & hours     &    $V$   & $R_{c}$  & $I_{c}$  \\
\hline
2011 Sep 30    &     9.4    &    145    &    145    &    145   &    2015 Jul 09    &    4.9    &    51    &    51    &    51    &    2015 Jun 07    &    4.6    &    54    &    57    &    57 \\
2011 Oct 01    &    10.1    &    156    &    156    &    156   &    2015 Jul 10    &    2.3    &    26    &    25    &    26    &    2015 Jun 09    &    4.8    &    61    &    62    &    55 \\
2011 Oct 15    &     9.8    &    153    &    153    &    153   &    2015 Jul 11    &    4.6    &    48    &    42    &    43    &    2015 Jun 10    &    5.2    &    57    &    58    &    57 \\
2011 Oct 16    &     5.8    &    92     &    92     &    92    &    2015 Jul 19    &    5.6    &    60    &    60    &    60    &    2015 Jun 11    &    5.2    &    59    &    56    &    61 \\
2011 Oct 21    &    10.9    &    165    &    165    &    165   &    2015 Jul 25    &    6.2    &    67    &    66    &    65    &    2015 Jun 13    &    5.3    &    62    &    68    &    71 \\
2011 Oct 22    &    11.4    &    173    &    173    &    173   &    2015 Aug 01    &    6.3    &    68    &    68    &    68    &    2015 Jun 16    &    2.7    &    35    &    38    &    38 \\
2011 Oct 23    &    11.2    &    170    &    170    &    170   &    2015 Aug 05    &    4.3    &    46    &    45    &    46    &    2015 Jun 25    &    3.6    &    47    &    47    &    47 \\
2018 Jan 31    &     4.0    &    50     &    50     &    50    &    2015 Sep 01    &    6.2    &    65    &    61    &    60    &    2015 Jun 30    &    4.4    &    62    &    62    &    62 \\
2018 Feb 05    &     5.5    &    68     &    68     &    68    &    2015 Sep 20    &    4.5    &    47    &    47    &    47    &    2015 Jul 01    &    3.8    &    49    &    54    &    54 \\
2018 Feb 07    &     2.1    &    27     &    27     &    27    &    2017 Sep 23    &    4.2    &    55    &    55    &    55    &    2015 Jul 06    &    4.8    &    66    &    66    &    66 \\
2018 Feb 12    &     2.7    &    34     &    34     &    34    &    2018 Jul 23    &    5.8    &    73    &    73    &    73    &    2019 May 11    &    5.4    &    66    &    66    &    66 \\
2018 Feb 13    &     4.0    &    50     &    50     &    50    &    2018 Aug 01    &    3.7    &    47    &    47    &    47    &                   &           &          &          &       \\
2019 Jan 20    &     5.8    &    71     &    71     &    71    &    2019 Aug 25    &    6.4    &    80    &    80    &    80    &                   &           &          &          &       \\
               &            &           &           &          &    2019 Aug 26    &    4.9    &    61    &    61    &    61    &                   &           &          &          &       \\
               &            &           &           &          &    2019 Aug 29    &    6.2    &    77    &    74    &    77    &                   &           &          &          &       \\
\hline
\end{tabular}
\end{table*}

\begin{table*}
\caption{The coordinates, $V$ magnitude and colours of the target stars, comparison stars and check star.}
\label{tab:tabCoo}
\begin{tabular}{llccccc}
\hline
       & star            &   $\alpha (2000.0)$                   &  $\delta (2000.0)$                &    $V$            & $V-R_{c}$        &  $V-I_{c}$      \\
\hline
target & TYC 3700-1384-1 &   $02^{\rmn{h}} 42^{\rmn{m}} 45\fs 3$ &  $+52\degr 59\arcmin 19\farcs 9$  &  $11.36 \pm 0.01$ &  $0.46 \pm 0.01$ & $0.90 \pm 0.02$  \\
comp 1 & TYC 3700-0950-1 &   $02^{\rmn{h}} 45^{\rmn{m}} 38\fs 5$ &  $+53\degr 02\arcmin 50\farcs 6$  &  $10.56 \pm 0.04$ &  $0.22 \pm 0.15$ & $0.42 \pm 0.15$  \\
comp 2 & GSC 3700-1333   &   $02^{\rmn{h}} 45^{\rmn{m}} 24\fs 0$ &  $+53\degr 00\arcmin 27\farcs 6$  &  $11.99 \pm 0.04$ &  $0.75 \pm 0.16$ & $1.45 \pm 0.16$  \\
comp 3 & GSC 3700-1121   &   $02^{\rmn{h}} 45^{\rmn{m}} 25\fs 5$ &  $+52\degr 57\arcmin 12\farcs 5$  &  $12.64 \pm 0.04$ &  $0.94 \pm 0.14$ & $1.82 \pm 0.14$  \\  
check  & TYC 3700-1332-1 &   $02^{\rmn{h}} 45^{\rmn{m}} 31\fs 9$ &  $+52\degr 59\arcmin 40\farcs 6$  &  $11.48 \pm 0.07$ &  $0.81 \pm 0.07$ & $1.56 \pm 0.07$  \\
       &                 &                                       &                                   &                   &                  &                  \\
target & V1511 Her       &   $17^{\rmn{h}} 55^{\rmn{m}} 27\fs 4$ &  $+44\degr 06\arcmin 54\farcs 4$  &  $11.51 \pm 0.01$ &  $0.27 \pm 0.02$ & $0.51 \pm 0.03$  \\
comp 1 & TYC 3101-687-1  &   $17^{\rmn{h}} 55^{\rmn{m}} 46\fs 3$ &  $+44\degr 08\arcmin 45\farcs 8$  &  $11.16 \pm 0.01$ &  $0.26 \pm 0.03$ & $0.50 \pm 0.03$  \\
comp 2 & TYC 3101-1015-1 &   $17^{\rmn{h}} 55^{\rmn{m}} 16\fs 3$ &  $+44\degr 01\arcmin 19\farcs 1$  &  $11.43 \pm 0.02$ &  $0.40 \pm 0.06$ & $0.77 \pm 0.06$  \\
comp 3 & GSC 3101-1069   &   $17^{\rmn{h}} 55^{\rmn{m}} 47\fs 0$ &  $+44\degr 11\arcmin 02\farcs 1$  &  $12.42 \pm 0.00$ &  $0.28 \pm 0.06$ & $0.54 \pm 0.06$  \\
comp 4 & GSC 3101-0805   &   $17^{\rmn{h}} 56^{\rmn{m}} 01\fs 3$ &  $+44\degr 02\arcmin 07\farcs 0$  &  $12.62 \pm 0.01$ &  $0.28 \pm 0.04$ & $0.54 \pm 0.04$  \\
check  & TYC 3101-1607-1 &   $17^{\rmn{h}} 55^{\rmn{m}} 28\fs 2$ &  $+44\degr 02\arcmin 23\farcs 2$  &  $13.32 \pm 0.01$ &  $0.24 \pm 0.08$ & $0.48 \pm 0.08$  \\
       &                 &                                       &                                   &                   &                  &                  \\
target & V1179 Her       &   $16^{\rmn{h}} 27^{\rmn{m}} 44\fs 9$ &  $+11\degr 03\arcmin 38\farcs 0$  &  $11.42 \pm 0.01$ &  $0.38 \pm 0.02$ & $0.72 \pm 0.04$  \\
comp 1 & TYC 963-108-1   &   $16^{\rmn{h}} 27^{\rmn{m}} 45\fs 5$ &  $+11\degr 08\arcmin 46\farcs 3$  &  $11.14 \pm 0.04$ &  $0.43 \pm 0.15$ & $0.83 \pm 0.15$  \\
comp 2 & TYC 963-266-1   &   $16^{\rmn{h}} 28^{\rmn{m}} 01\fs 7$ &  $+11\degr 08\arcmin 17\farcs 7$  &  $12.11 \pm 0.02$ &  $0.62 \pm 0.16$ & $1.20 \pm 0.16$  \\
comp 3 & GSC 963-96      &   $16^{\rmn{h}} 27^{\rmn{m}} 38\fs 1$ &  $+11\degr 10\arcmin 36\farcs 5$  &  $12.03 \pm 0.04$ &  $0.40 \pm 0.16$ & $0.78 \pm 0.16$  \\
check  & GSC 964-69      &   $16^{\rmn{h}} 28^{\rmn{m}} 10\fs 7$ &  $+11\degr 03\arcmin 32\farcs 5$  &  $13.02 \pm 0.02$ &  $0.44 \pm 0.18$ & $0.86 \pm 0.16$  \\
\hline
\end{tabular}
\begin{flushleft}
Note: For the target stars the magnitude and colours at maximum light are listed, obtained from the observations presented in this paper. The magnitude and colours
for the comparison and check stars are taken from the AAVSO Photometric All Sky Survey (APASS). Refer to the text for the discussion on the low precission of the
comparison stars colour indices.
\end{flushleft}
\end{table*}

\begin{table*}
\caption{TYC 3700-1384-1 comparison star matrix. \label{tab:compmatrix}}
\begin{tabular}{ccccc}
\hline
            & \multicolumn{4}{|c|} {measured }                                                 \\
\cline{2-5}			
            & comp 1             & comp 2             & comp 3             & check               \\
\hline
comp        & \multicolumn{4}{|c|} {$V$}  \\
comp 1      &                    & $11.990 \pm 0.008$ & $12.619 \pm 0.018$ & $11.470 \pm 0.009$  \\
comp 2      & $10.554 \pm 0.008$ &                    & $12.615 \pm 0.018$ & $11.466 \pm 0.010$  \\
comp 3      & $10.582 \pm 0.018$ & $12.013 \pm 0.018$ &                    & $11.494 \pm 0.014$  \\
\hline
comp & \multicolumn{4}{|c|} {$V-R_{c}$}  \\
comp 1      &                   & $0.759 \pm 0.009$ & $0.940 \pm 0.014$ & $0.787 \pm 0.008$  \\
comp 2      & $0.208 \pm 0.010$ &                   & $0.932 \pm 0.015$ & $0.779 \pm 0.011$  \\
comp 3      & $0.218 \pm 0.014$ & $0.761 \pm 0.016$ &                   & $0.789 \pm 0.014$  \\
\hline
comp & \multicolumn{4}{|c|} {$V-I_{c}$} \\
comp 1      &                   & $1.546 \pm 0.010$ & $1.797 \pm 0.014$ & $1.524 \pm 0.008$ \\
comp 2      & $0.414 \pm 0.010$ &                   & $1.789 \pm 0.014$ & $1.515 \pm 0.010$ \\
comp 3      & $0.440 \pm 0.014$ & $1.474 \pm 0.015$ &                   & $1.542 \pm 0.014$ \\
\hline
\end{tabular}
\end{table*}

\section{Light curve analysis}
\begin{figure}
\includegraphics[width=\columnwidth]{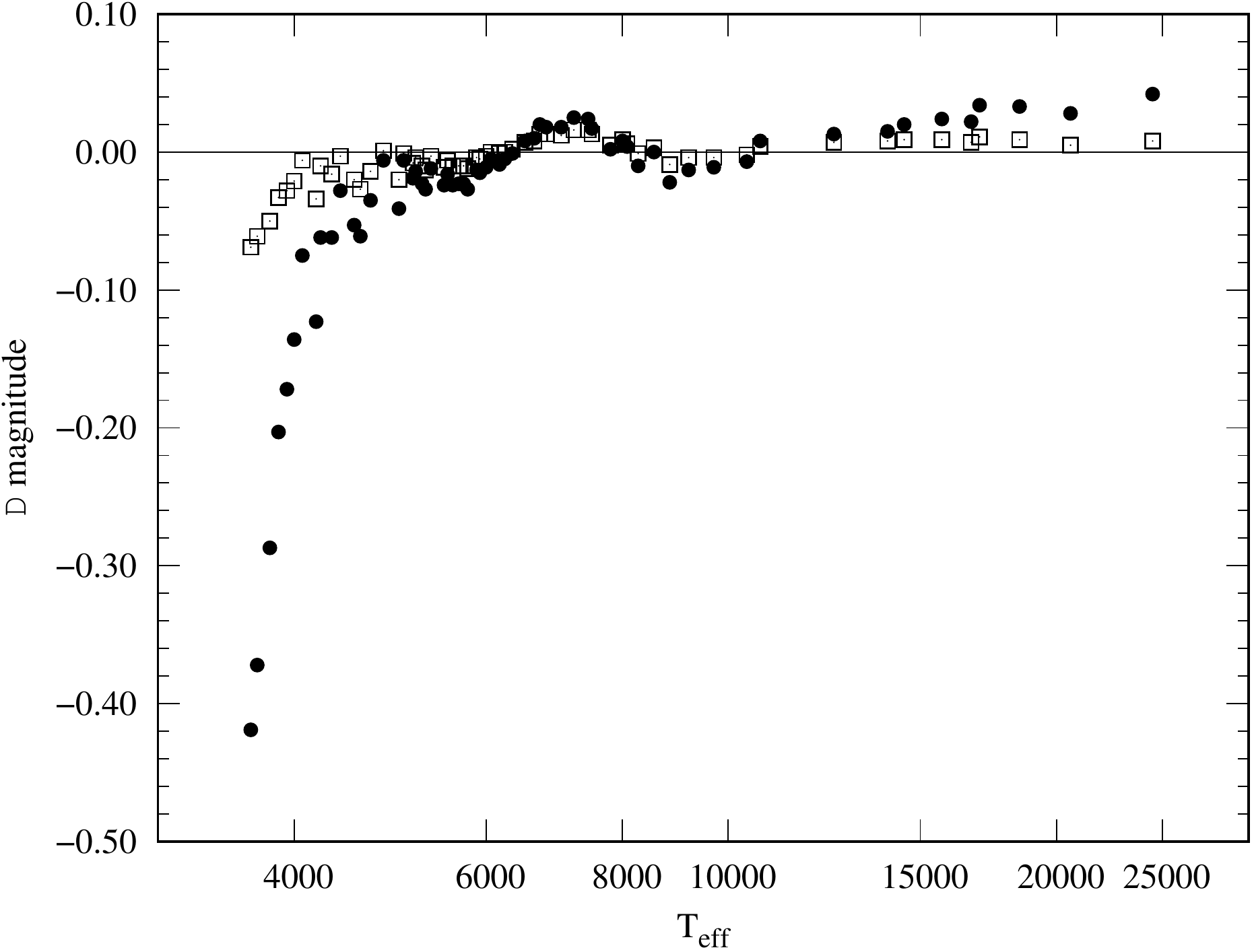}
\caption{
Difference of the $V-R_{c}$ (open squares) and $V-I{c}$ (filled circles) colour indices obtained from the \citet{castelli03} responses for a single dwarf star in {\sc phoebe 2}, minus
the values provided in the online table of \citet{pecaut13} in function of {\Teff}.  
}
\label{fig:CI_comp}
\end{figure}

The light curves were initially analysed using the Wilson-Devinney (WD) code \citep{wilson71,wilson72,wilson79,wilson94} implemented in the legacy version of {\sc phoebe} \citep{prsa05}.
The effective temperature of the more massive star, the bolometric albedos and the gravity-darkening coefficients were fixed. As will be shown in following subsections, all three stars are
of spectral class F, hence the gravity-darkening coefficients have been fixed to $g_1=g_2=0.32$ \citep{lucy67} and the albedos to $A_1=A_2=0.5$ \citep{rucinski69} appropriate for stars
with a convective envelope. The logarithmic limb-darkening law coefficients from \citet{vanhamme93} for a solar composition star were updated automatically during the fitting process.
The modelling was performed for the $VR_{c}I_{c}$ light curves simultaneously in mode 3, which is appropriate for overcontact
systems that are not in thermal contact. The adjustable parameters were: the orbital inclination $i$; the mean temperature of star 2, $T_2$;
the monochromatic luminosities of star 1, $L_1$ in each observed passband; and the dimensionless potential of star 1, $\Omega_1$ ($\Omega_1 = \Omega_2$ for contact binaries, $\Omega$ will be used
further in this paper).
It is known that the formal errors provided by the WD code are unrealistically small. 
The anonymous referee pointed out that the approach of assuming a fixed temperature for one component is not adequate, and that the method outlined in \citet{prsa05} is capable
to provide individual temperatures and eventually should result in the right object colour. Additionally, the error distributions should be sampled to provide reliable parameter uncertainties.
Since the {\sc phoebe 2} versions \citep{prsa16, horvat18, jones20, conroy20} provide the feature for colour-constraining the model passband fluxes, and a Markov Chain Monte Carlo (MCMC) sampler based on
{\sc emcee} \citep{foreman-mackey13}, the analysis has been re-done using {\sc phoebe 2}. The provisional values obtained with the legacy version of {\sc phoebe} are listed in Table \ref{tabPhotSol} and 
served as priors for the MCMC sampling. In the last row of the table the reduced $\chi^2_{r}$ is provided as the sum of the reduced $\chi^2_{r}$ values for the 3 passbands separately. 
The observed magnitudes were transformed to fluxes using a passband-independent magnitude $m0$.
The value for $m0$ was chosen in a way that the flux of the $V$ light curve is of the order of unity at first quadrature phase and is given in Table \ref{tab:tabCoo}. 
Next, a synthetic single A0V (${\Teff} = 9600 K$) star was created in {\sc phoebe 2} and the passband flux ratios $f_{V}/f_{Rc}$ and $f_{V}/f_{Ic}$,
based on the \citet{castelli03} model atmospheres, were calculated to scale the observed fluxes to the \citet{castelli03} responses in {\sc phoebe 2}.
In order to test the validity of this approach, colour indices were calculated for synthetic single stars in the range of $3550 K \lid {\Teff} \lid 26 000 K$ and compared with the colour indices given
in the \citet{pecaut13}\footnote{\url{https://www.pas.rochester.edu/~emamajek/EEM\_dwarf\_UBVIJHK\_colors\_Teff.txt}} online table, version 2019.3.22.
With the exception of the extreme cooler and hotter ends of this range, the differences of the calculated colour indices are not larger than $2\%$ as shown in Fig. \ref{fig:CI_comp}.
Since the studied stars have been observed in 3 passbands the temperatures are overdetermined. Therefore for each star two MCMC runs were executed, one for the combination of the $V$ and $R_{c}$
passband and one for the $V$ and $I_{c}$ passband.

\subsection{TYC 3700-1384-1} \label{subsection:TYC 3700-1384-1}
\begin{figure}
\includegraphics[width=\columnwidth]{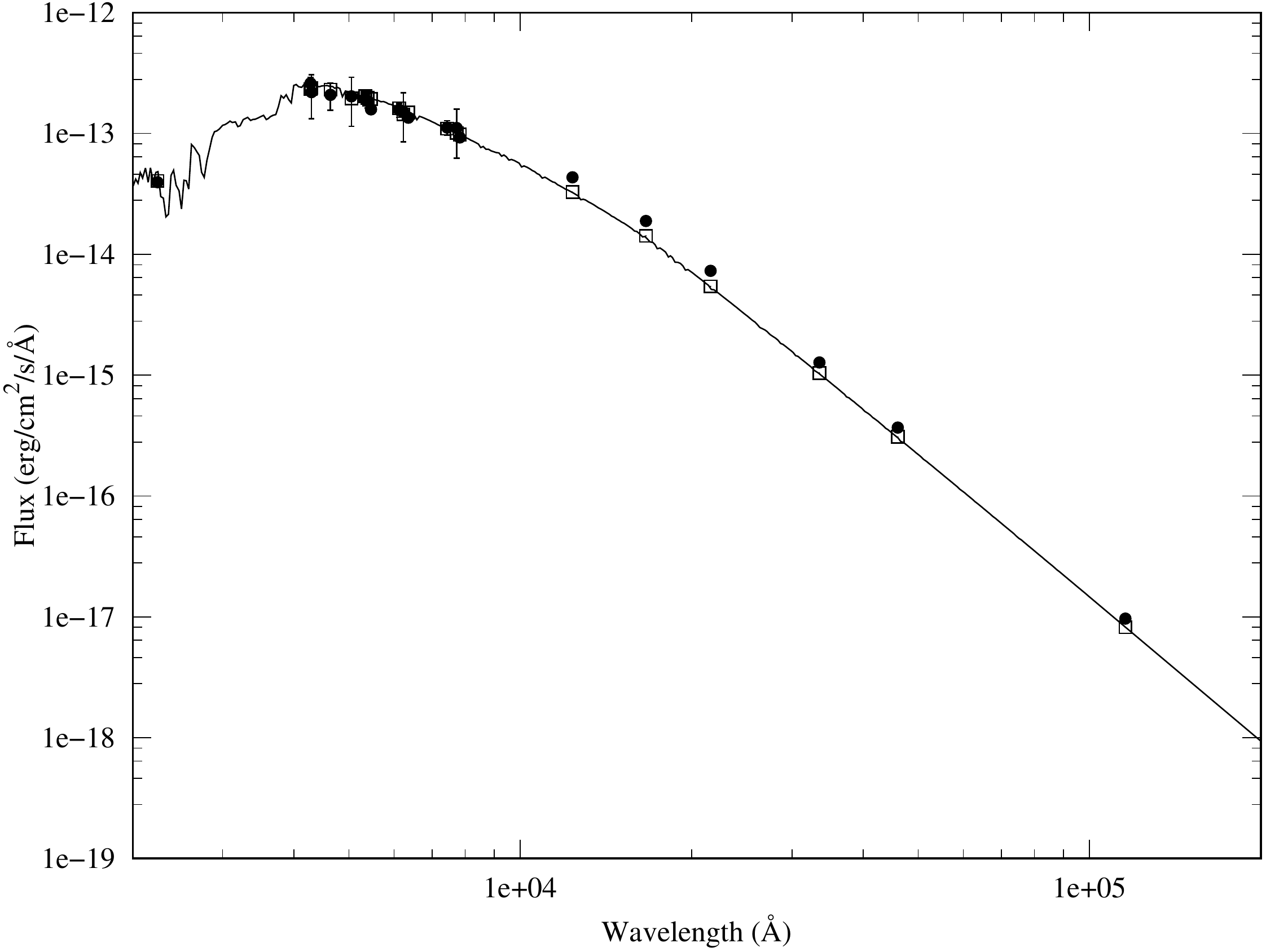}
\caption{
SED of TYC 3700-1384-1, the solid line is the Kurucz ODFNEW/NOVER model for a star with ${\Teff} = 6750K$, $log g~=~4.0$ and solar metalicity.
The solid points are the fluxes calculated from the publicly available GALEX, APASS, SDSS, GAIA, 2MASS and WISE photometry, together with the $VR_{c}I_{c}$ observations
at secondary minimum presented in this paper.}
\label{fig:SED}
\end{figure}

\begin{figure}
\includegraphics[width=\columnwidth]{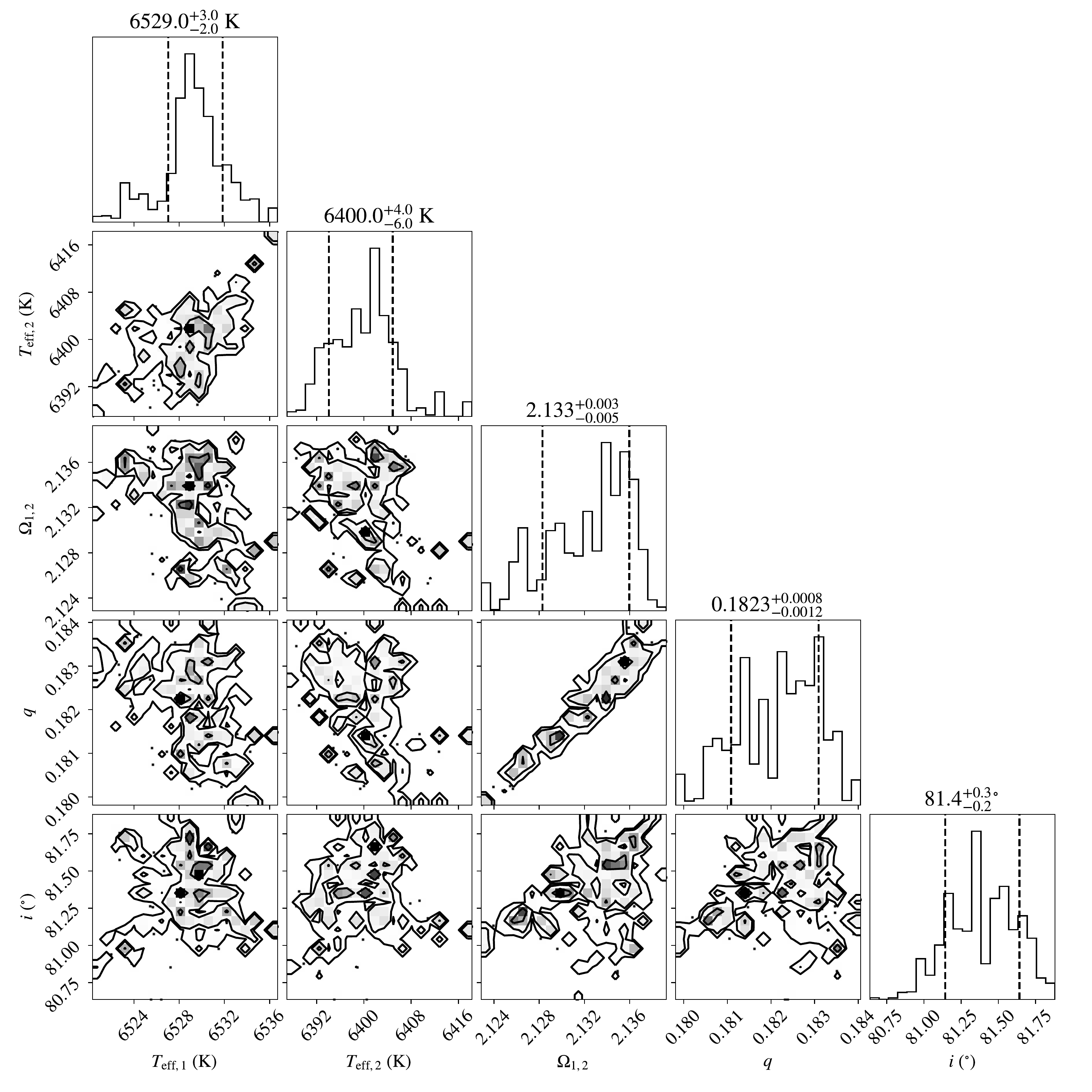}
\caption{
Corner plot depicting the {\sc phoebe 2} MCMC posterior distributions for the effective temperatures of the primary ($T_{1}$) and secondary ($T_{2}$), the dimensionless Kopal potential ($\Omega_{1,2}$),
the mass ratio ($q$) and the inclination ($i$) from the run with the $V$ and $R_{c}$ passbands of TYC 3700-1384-1. The dashed lines indicate the 16th and 84th percentile.
}
\label{fig:TYC3700-1384-1_corner_VR}
\end{figure}

\begin{figure}
\includegraphics[width=\columnwidth]{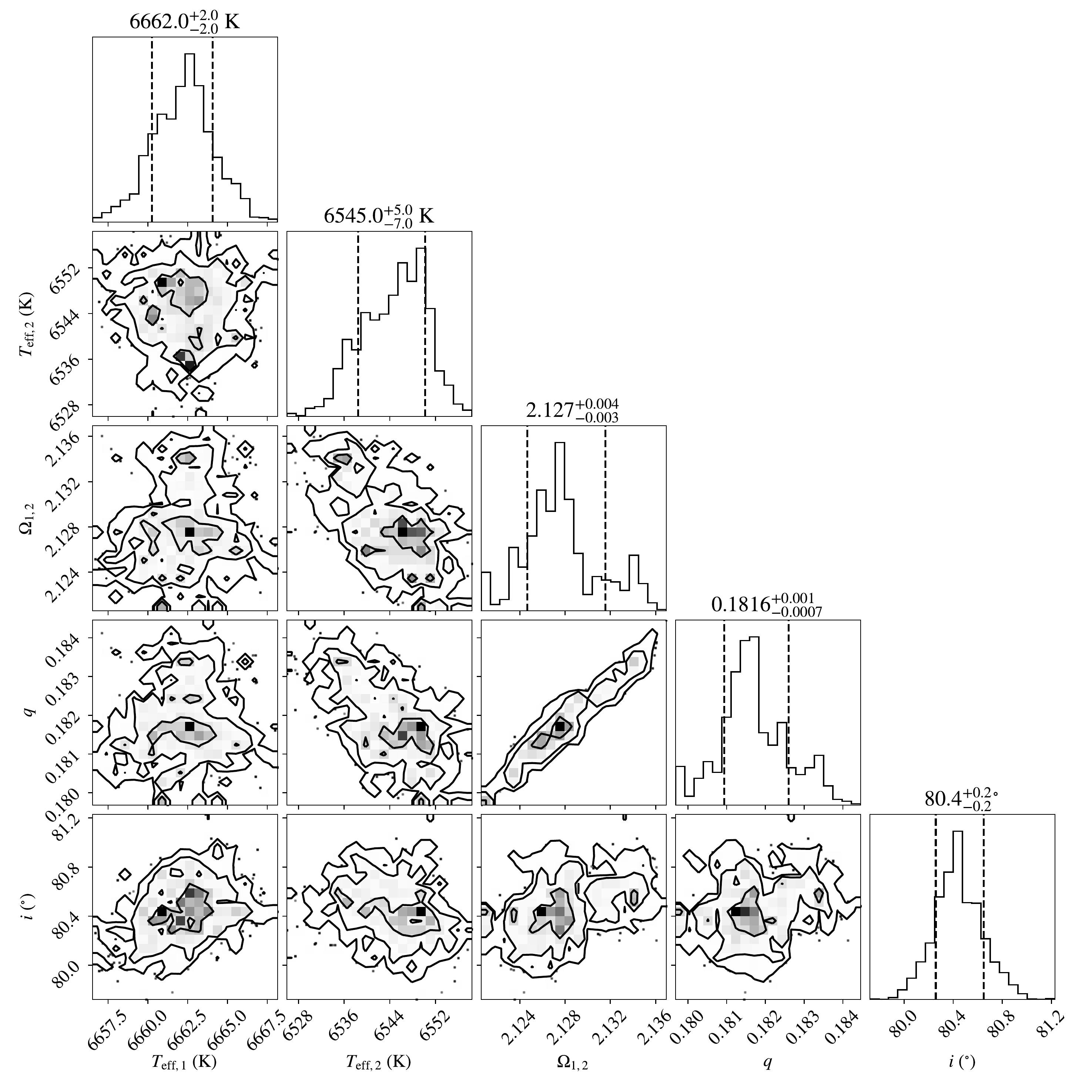}
\caption{
Same as Fig. \ref{fig:TYC3700-1384-1_corner_VR} but from the run with the $V$ and $I_{c}$ passbands of TYC 3700-1384-1.
}
\label{fig:TYC3700-1384-1_corner_VI}
\end{figure}

\begin{figure}
\includegraphics[width=\columnwidth]{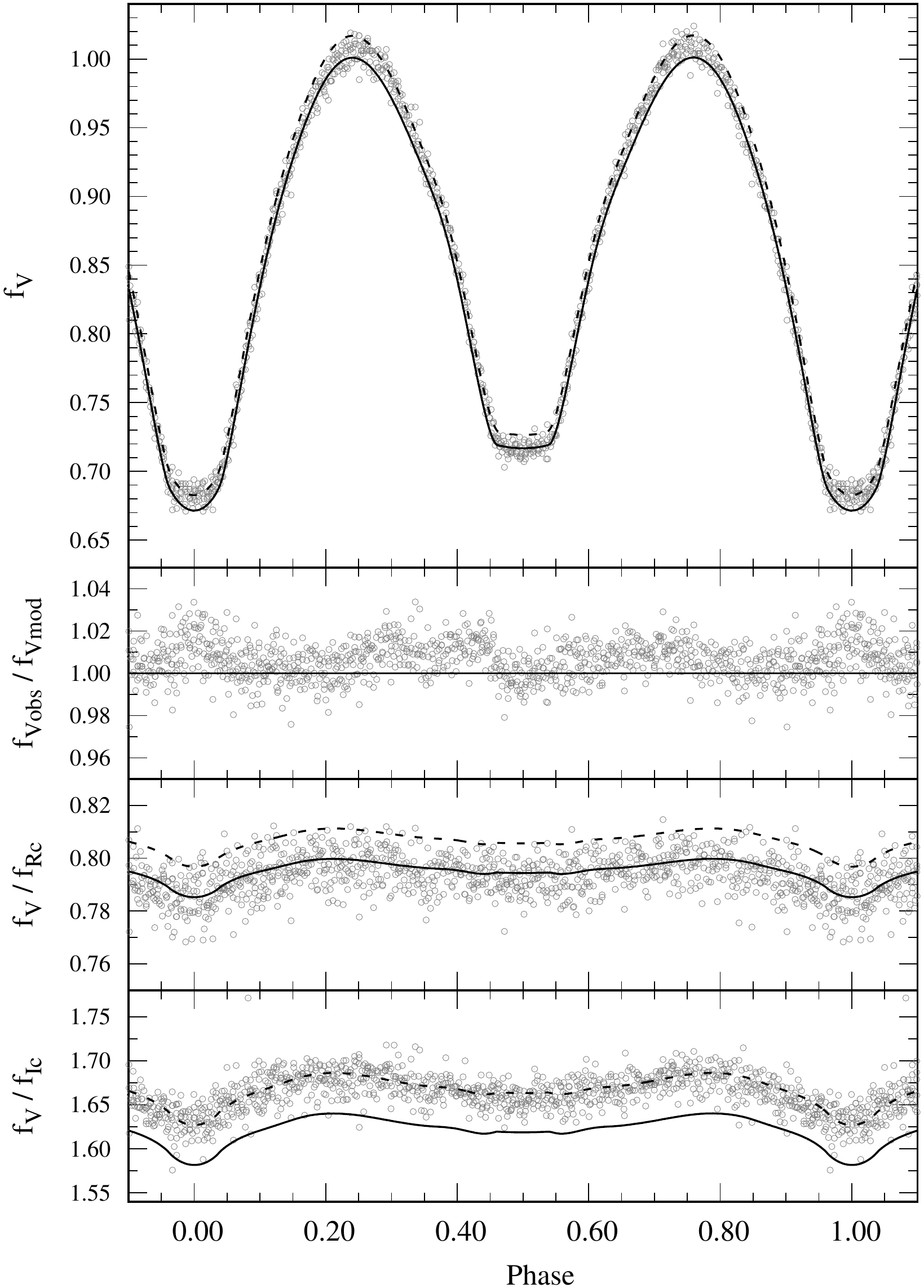}
\caption{
$V$ light curve of TYC 3700-1384-1 as normalized flux. The solid line is the synthetical light curve based on the results of the MCMC run with the $VR_{c}$ observations and the dashed
line on the run with the $VI_{c}$ observations. 
The lower panels display the residuals as the flux ratio of the $V$ observations to the $VR_{c}$ model, and the colour curves respectively.}
\label{fig:lc_TYC3700}
\end{figure}

For the initial analysis with the legacy version of {\sc phoebe}, the effective temperature of star 1, $T_1$, was estimated by fitting the spectral
energy distribution (SED) to Kurucz ODFNEW/NOVER models using VOSA\footnote{http://svo2.cab.inta-csic.es/theory/vosa/} \citep{bayo08} .
The best model is selected through $\chi^2$ minimization. Publicly available broad-band photometry from UV to IR wavelenghts, as well as the photometry
from the observations presented in this paper, have been used to construct the SED. 
To fit the SED, the surface gravity was fixed to $log g = 4$ and the metallicity is fixed to be solar, which is justified by the study from
\citet{rucinski13} showing the metallicity to be roughly solar from a sample of 52 F-type contact binaries. The SED is corrected for
interstellar reddening with $E(B-V) = 0.273$ based on the reddening $E(g-r)_{PS1} = 0.279$ in the Pan-STARRS 1 passbands, provided by
the Bayestar19 dustmap \citep{green19} for a $Gaia$ DR2 distance of $335 \pm 4 pc$ \citep{gaia18}. 
The reddening $E(g-r)_{PS1}$ is converted to $E(B-V)$ using $E(B-V) = 0.981 E(g-r)_{PS1}$ as derived from table 6 in \citet{schlafly11}.
The distance is however slightly smaller than the minimum reliable distance modulus $m-M = 8.26$ indicated for this line-of-sight. 
The best fit was found for a ${\Teff} = 6750 \pm 125 K$ and is plotted in Fig. \ref{fig:SED}.
Fixing $T_1$ to this value, the photometric solution converges to a low mass ratio of $q=0.19$, an inclination $i=82\fdg1$, a slightly cooler secondary with
${\Teff}= 6576 K$, and $\Omega = 2.15$.
These values were used as priors, with reasonable uniform distributions, for the MCMC parameter space sampling. 
The observed fluxes were de-reddened using table 3 in \citet{cardelli89} and the effective temperature of the primary was added to the set
of parameters allowed to vary freely. A corner plot of the resulting MCMC chain for the set of freely varying parameters is shown for the
$VR_{c}$ run in Fig. \ref{fig:TYC3700-1384-1_corner_VR} and for the $VI_{c}$ run in Fig. \ref{fig:TYC3700-1384-1_corner_VI}.
The model parameters, as extracted from the posterior distributions are listed in Table \ref{tab:mmcposteriors}. The last row lists
the average of the obtained values with the standard deviation as uncertainty estimate. For the
temperatures the largest internal errors of each component as obtained from the MCMC posterior distribution are added in quadrature. 
The effective temperatures for the primary and secondary component are $T_{1} = 6596 \pm 98K$ and $T_{2} = 6472 \pm 106 K$ 
respectively. The inclination of the system $i = 80.9 \pm 0.7 \degr$ shows that the eclipses are seen nearly edge-on.
The mass ratio $q = 0.182 \pm 0.001$ and $\Omega = 2.130 \pm 0.004$ reveal a moderate fill-out factor of $f = 49 \%$.
The light and colour curves are shown in Fig. \ref{fig:lc_TYC3700}. The solid line and dashed line are based on the results from the MCMC runs with the $VR_{c}$ and $VI_{c}$ observations
respectively. The residuals of the $V$ light curve are plotted as the flux ratio to the model from the $VR_{c}$ run.
Because ${\Teff}$ is overdetermined with more than 2 bandpasses, one of the model's colour curves shows a mismatch in flux level with respect to the observations. This is as expected,
unless in the unlikely case that all systematic errors are negligible small. The model reproduces the shape of the observed colour curves very well. \\
\citet{pasternacki11} derived very different stellar parameters for this star based on observations with the Berlin Exoplanet Search Telescope.
The results listed in their table 2 appear to be incorrect. They found a temperature difference of $\Delta{\Teff} = 670K$ between the primary and secondary star,
with the primary star the cooler one. However, the light curve published in their paper cannot account for such a temperature difference because the depth of the primary
and secondary eclipses are nearly equal. Also the values of the dimensionless Kopal potentials are too large.

\subsection{V1511 Her}
\begin{figure}
\includegraphics[width=\columnwidth]{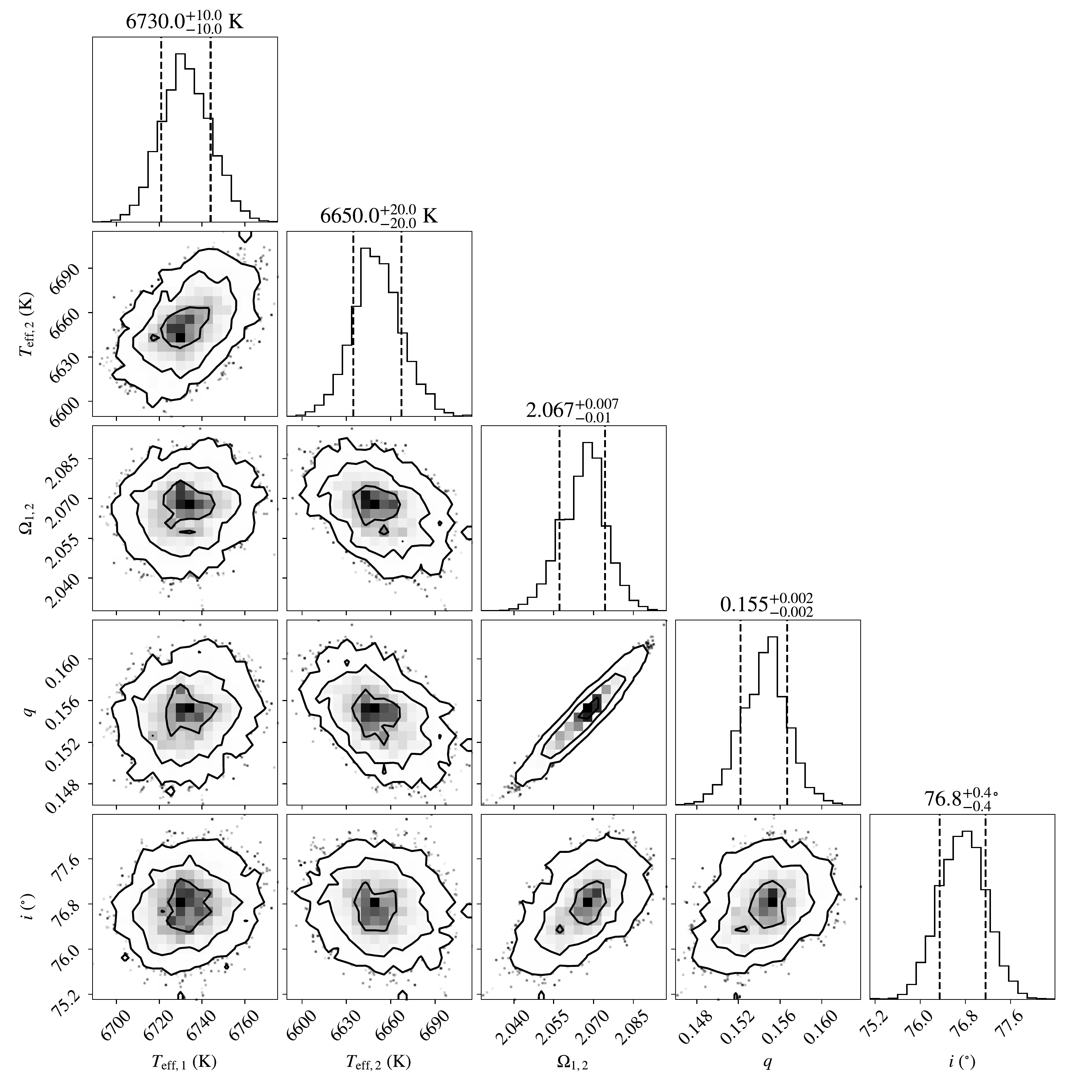}
\caption{
Corner plot depicting the {\sc phoebe 2} MCMC posterior distributions for the effective temperatures of the primary ($T_{1}$) and secondary ($T_{2}$), the dimensionless Kopal potential ($\Omega_{1,2}$),
the mass ratio ($q$) and the inclination ($i$) from the run with the $V$ and $R_{c}$ passbands of V1511 Her. The dashed lines indicate the 16th and 84th percentile.
}
\label{fig:V1511Her_corner_VR}
\end{figure}

\begin{figure}
\includegraphics[width=\columnwidth]{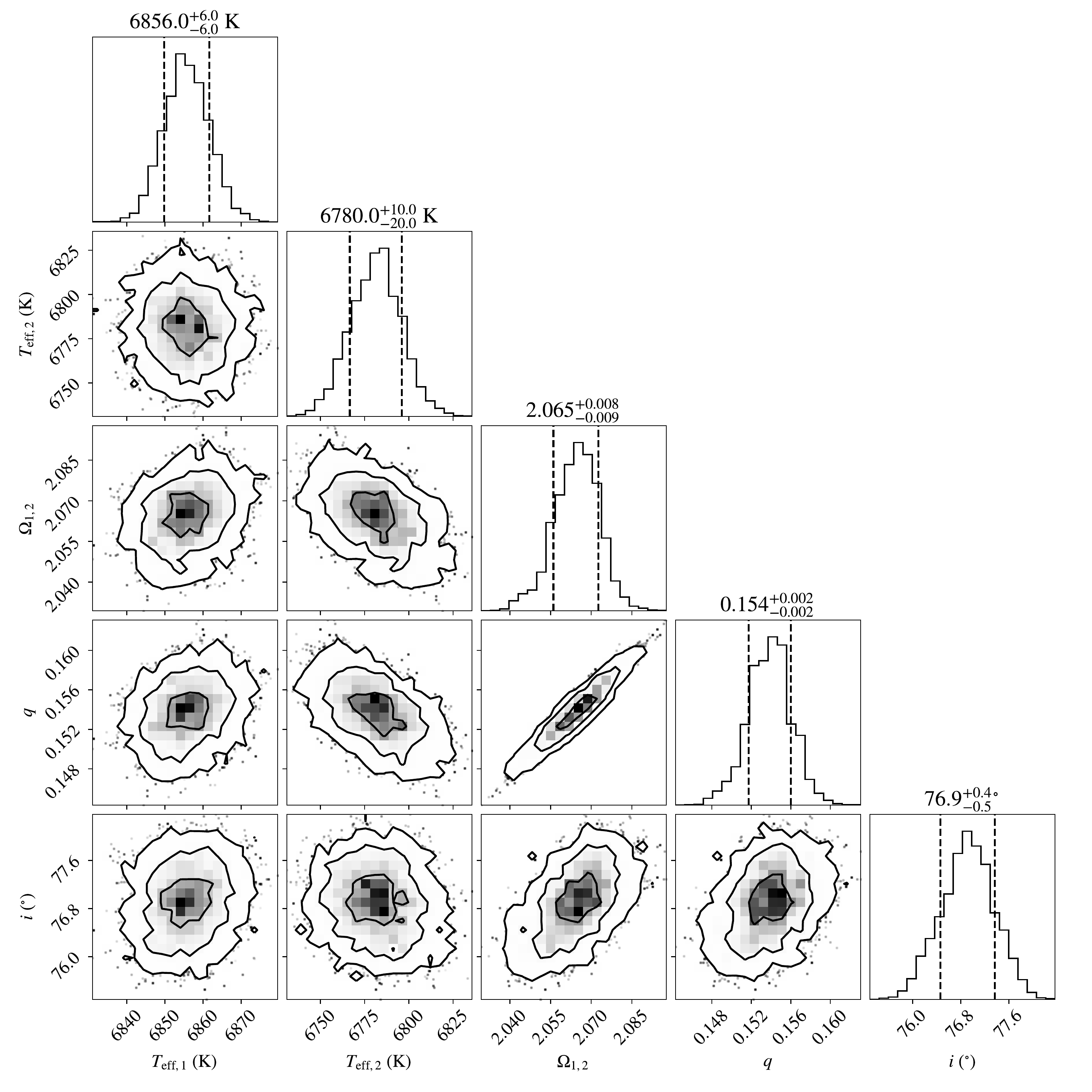}
\caption{
Same as Fig. \ref{fig:V1511Her_corner_VR} for the $V$ and $I_{c}$ passbands of V1511 Her.
}
\label{fig:V1511Her_corner_VI}
\end{figure}

\begin{figure}
\includegraphics[width=\columnwidth]{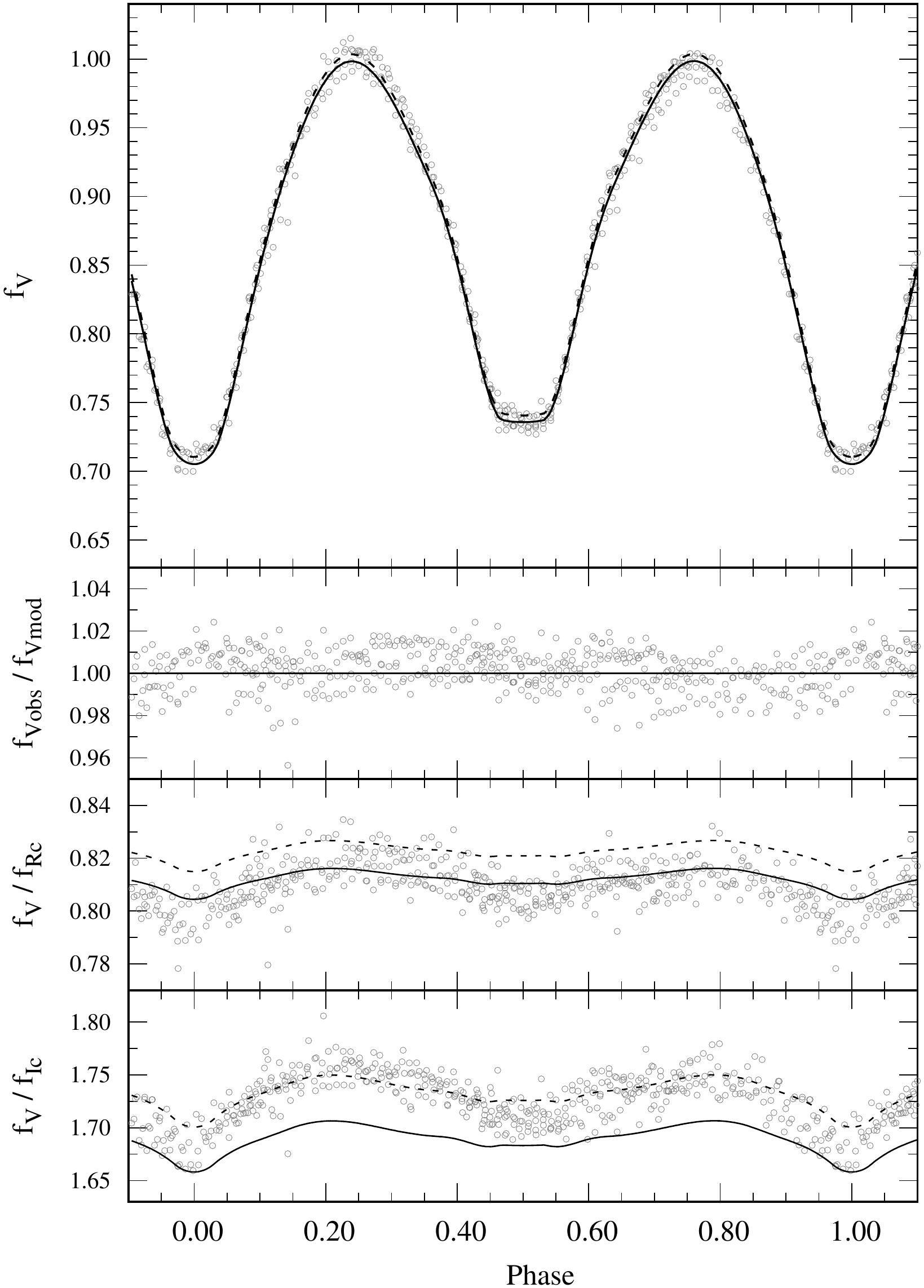}
\caption{
$V$ light curve of V1511 Her as normalized flux. The solid line is the synthetical light curve based on the results of the MCMC run with the $VR_{c}$ observations and the dashed
line on the run with the $VI_{c}$ observations. 
The lower panels display the residuals as the flux ratio of the $V$ observations to the $VR_{c}$ model, and the colour curves respectively.}
\label{fig:lc_V1511Her}
\end{figure}

V1511 Her is included in the fourth data release \citep[DR4\footnote{http://cdsarc.u-strasbg.fr/viz-bin/cat/V/153};][]{luo18} of the Large Sky Area
Multi-Object Fiber Spectroscopic Telescope \citep[LAMOST;][]{cui12,zhao12} regular survey.
The catalogue lists an effective temperature of ${\Teff} = 6096 \pm 17K$, a $log g = 4.2 \pm 0.03$ and a metallicity $[FE/H] = -0.04 \pm 0.01$.
The $V-I_{c}$ colour index from our observations indicate a much higher temperature. As discussed in section \ref{section:observations}, the APASS magnitudes
for this field might not be as accurate as the uncertainties indicate because the field was only observed on two nights.
$Gaia$ DR2 lists a distance of $315 \pm 19 pc$ for V1511 Her. For this distance, the Bayestar19 dust map gives a converted reddening of $E(B-V)=0.044$. 
De-reddening the $V-I_{c}$ colour index provided by \citet{terrell12}, $(V-I_{c})_0$ = 0.61, suggest a temperature that is in good agreement with the one provided by LAMOST DR4. However, the 
de-reddened $(B-V)_{0} = 0.45$ colour index suggest a much higher binary temperature of $\sim 6500 K$, while our reddening free $(V-I_{c})_0 = 0.44$ suggests a ${\Teff} \sim 6750 K$. 
Despite these large discrepancies, a light curve analyses has been performed. Since the star exhibits total eclipses and the 
important mass-ratio $q$ is not strongly correlated with the temperatures, the mass-ratio can be derived fairly accurately. 
Initially the light curve has been modelled using the legacy version of {\sc phoebe} with the temperature of the primary fixed to the value ${\Teff} = 6096 K$ provided by LAMOST.
The preliminary photometric solution resulted in a marginal temperature difference between the primary and the secondary, $T_2 = 6107 K$, and results
in a low mass ratio of $q=0.13$ and $\Omega = 1.984$, implying a very high fill-out factor of $f=80\%$. The inclination $i = 74\fdg 7$.
A photometric solution has also been obtained in a similar way as for TYC 3700-1384-1.
The parameter space was scanned using MCMC in {\sc phoebe 2}. The values for the priors were taken from the preliminary solution obtained with the legacy version,
except for the temperature of the primary. The prior for ${T_1}$ was set to $6760 K$ in agreement with the $V-I_{c}$ colour index of the observations.
Generous uniform distributions were used for all parameters.
The corner plots of the resulting MCMC chain are shown in Fig. \ref{fig:V1511Her_corner_VR} and Fig. \ref{fig:V1511Her_corner_VI} for the $VR_{c}$ and $VI_{c}$ run respectively.
The temperatures for the primary and secondary from this solution are ${T_1} = 6793 \pm 101K$ and ${T_1} = 6715 \pm 103K$. The significant difference in the colour indices with respect to
those published by \citet{terrell12} indicate that the systematic error of the component's effective temperatures is likely much larger than the uncertainties on these temperatures indicate.
The derived mass-ratio $q = 0.153 \pm 0.001$ in this solution is somewhat higher than the one obtained from the preliminary solution with the primary temperature fixed to the LAMOST DR4 binary temperature, 
while the inclination $i =  76\fdg 7 \pm 0\fdg2$ is similar. The value of $\Omega = 2.063 \pm 0.004$ results in a fill-out factor $f=49\%$ which is much smaller compared to the model
with the lower temperatures. The light curve and colour curves are plotted in Fig. \ref{fig:lc_V1511Her}. The solid line and dashed line are based on the results from the MCMC runs with
the $VR_{c}$ and $VI_{c}$ observations respectively. The observed colour variation is larger than calculated from the model. The amplitude of the $V-I_{c}$ colour variation in the preliminary
model based on the lower LAMOST DR4 temperature, is only $5 mmag$ larger, and hence does not account for the observed colour variation either.

\subsection{V1179 Her}
\begin{figure}
\includegraphics[width=\columnwidth]{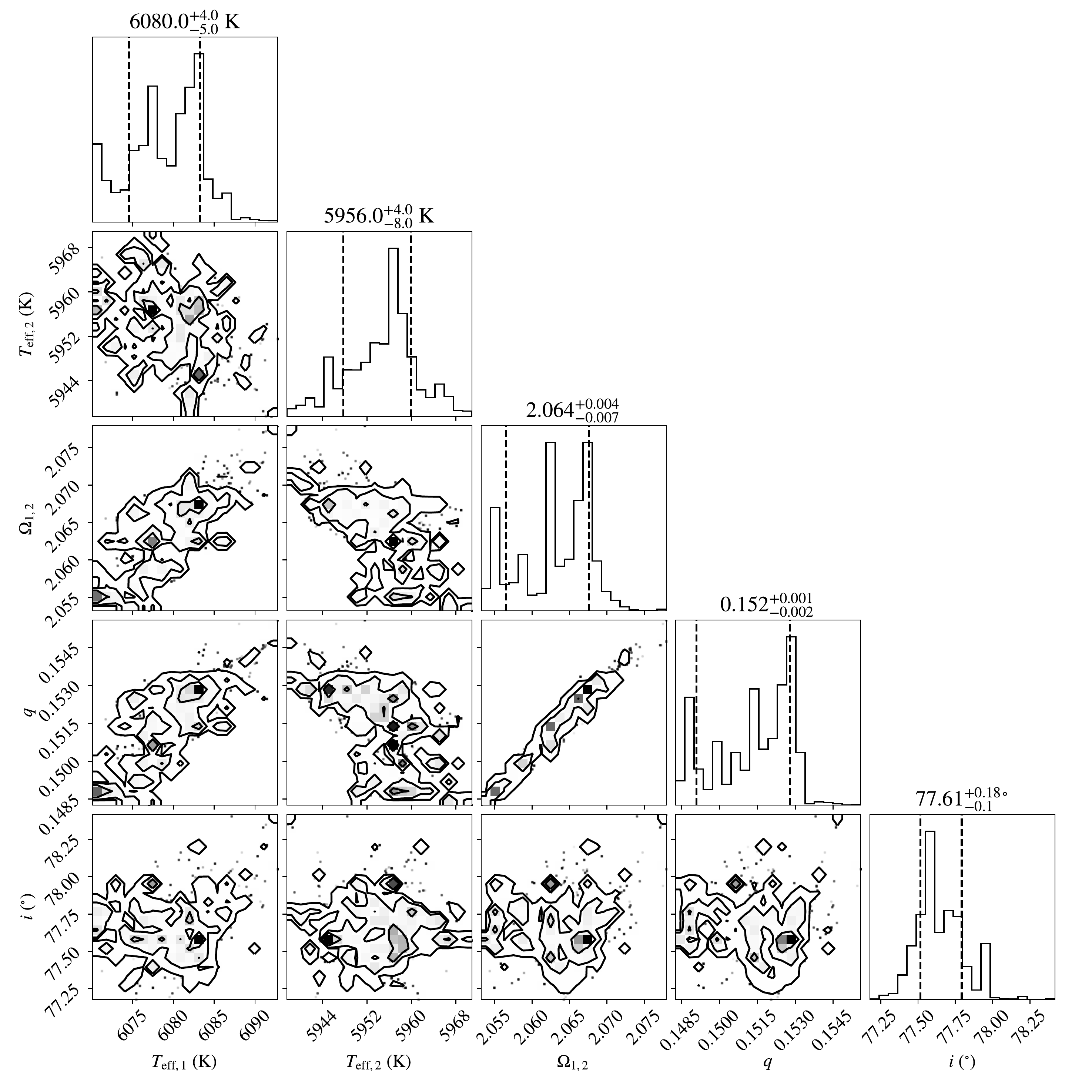}
\caption{
Corner plot depicting the {\sc phoebe 2} MCMC posterior distributions for the effective temperatures of the primary ($T_{1}$) and secondary ($T_{2}$), the dimensionless Kopal potential ($\Omega_{1,2}$),
the mass ratio ($q$) and the inclination ($i$) from the run with the $V$ and $R_{c}$ passbands of V1179 Her. The dashed lines indicate the 16th and 84th percentile.
}
\label{fig:V1179Her_corner_VR}
\end{figure}

\begin{figure}
\includegraphics[width=\columnwidth]{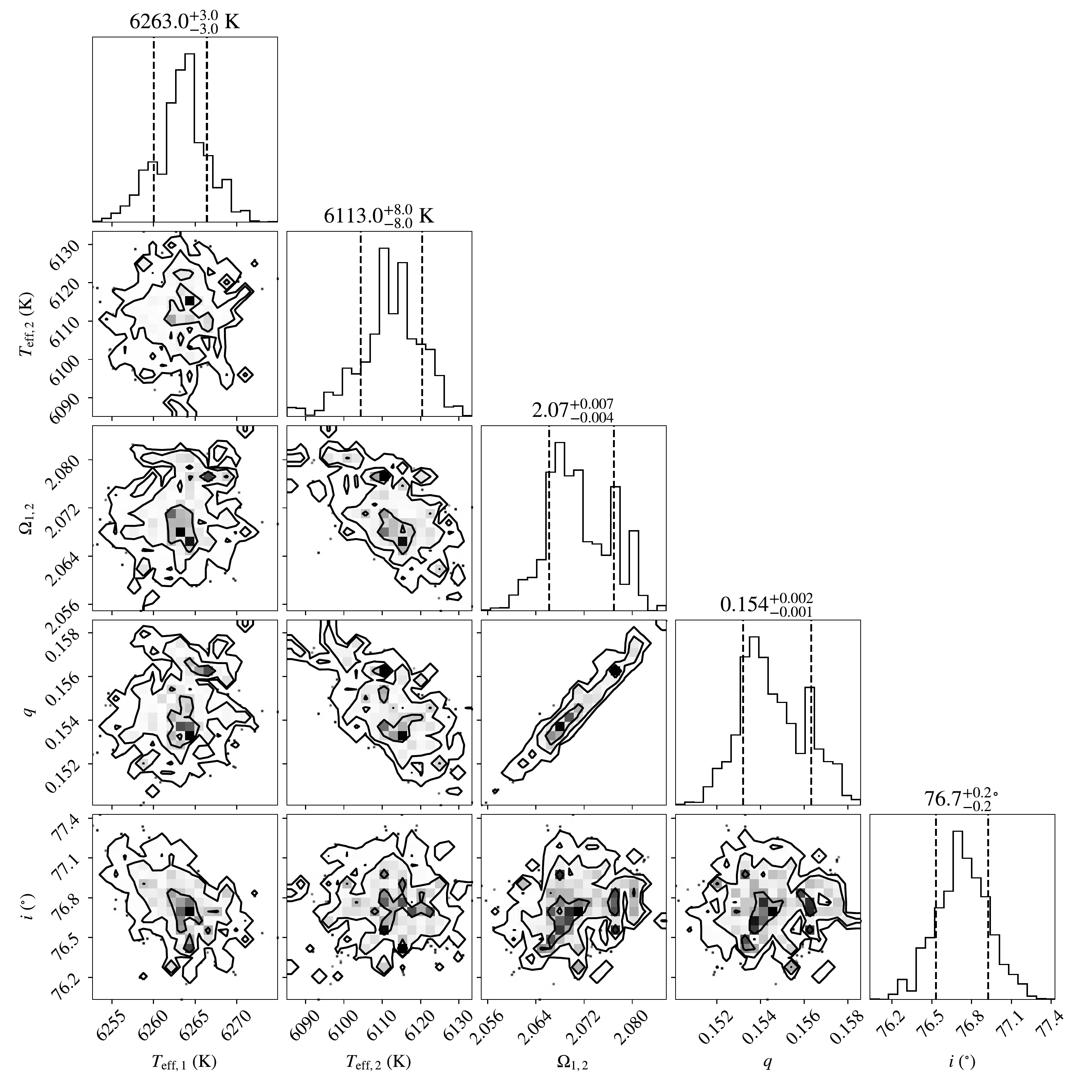}
\caption{
Same as Fig. \ref{fig:V1179Her_corner_VR} for the $V$ and $I_{c}$ passbands of V1179 Her.
}
\label{fig:V1179Her_corner_VI}
\end{figure}

\begin{figure}
\includegraphics[width=\columnwidth]{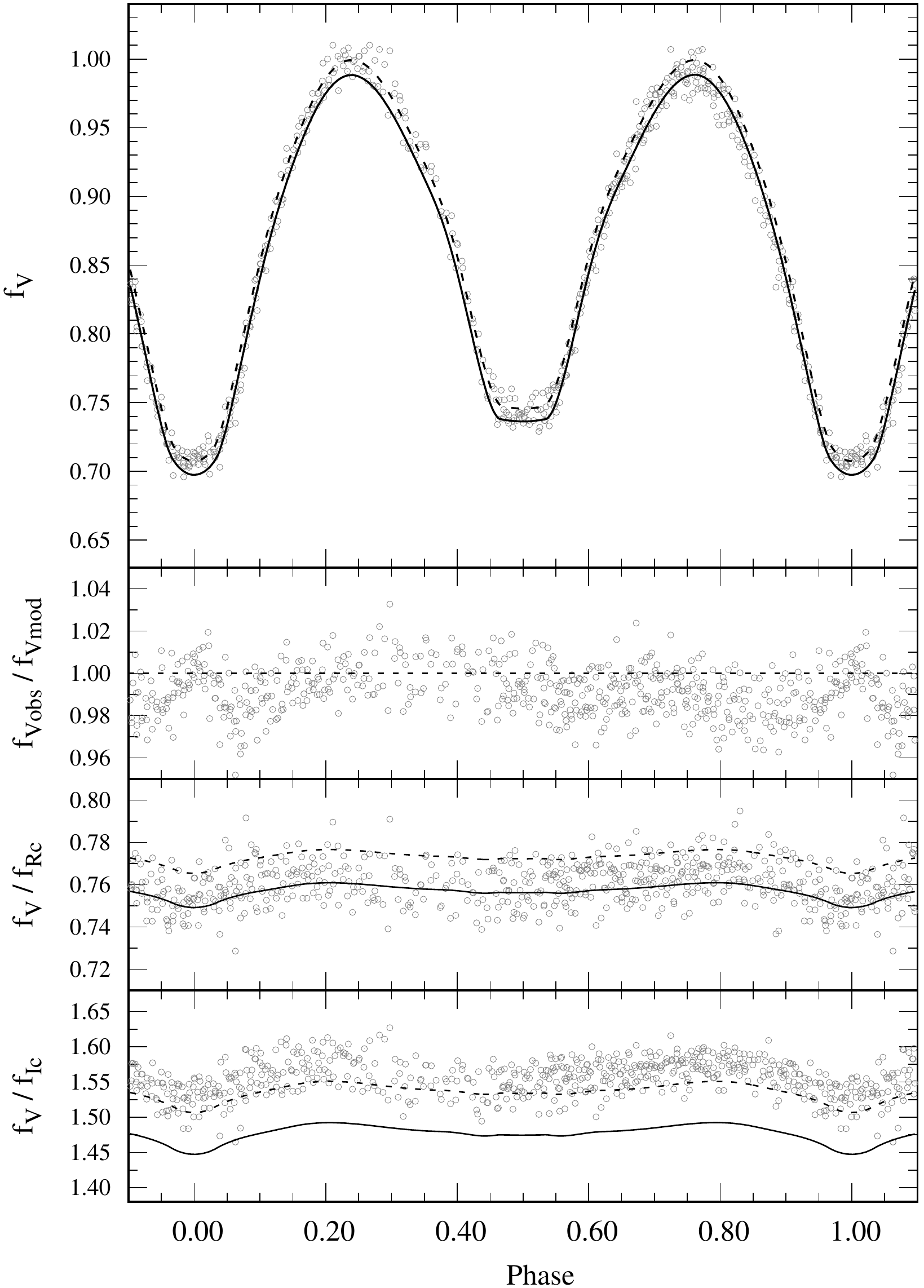}
\caption{
$V$ light curve of V1179 Her as normalized flux. The solid line is the synthetical light curve based on the results of the MCMC run with the $VR_{c}$ observations and the dashed
line on the run with the $VI_{c}$ observations. 
The lower panels display the residuals as the flux ratio of the $V$ observations to the $VI_{c}$ model, and the colour curves respectively.}
\label{fig:lc_V1179Her}
\end{figure}

LAMOST DR4 lists an effective temperature of ${\Teff} = 6249 \pm 12 K$,
a $log g = 4.12 \pm 0.02$ and a metallicity $[FE/H] = -0.12 \pm 0.01$, hence the temperature of star 1 was fixed to $T_1 = 6249 K$ for the initial analysis.
The best photometric solution for this provisional analysis is found for $T_{2}=6085 K$, a mass-ratio $q = 0.14$, $\Omega = 2.02$ and an inclination $i=77\fdg1$.
These parameters were used as priors for the parameter space sampling with MCMC. The analysis was performed in the same way as for the other two studied stars.
Using the reddening of $E(B-V) = 0.106$, provided by the Bayestar19 dust map for a distance of $403 \pm 5 pc$ as listed by $Gaia$ DR2, the observations
were de-reddened by applying the passband dependent ratios given in table 3 in \citet{cardelli89}.
The results from the two individual MCMC runs in {\sc phoebe 2} are summarized in table \ref{tab:mmcposteriors}, the corner plots of the resulting MCMC chains are
shown in Fig. \ref{fig:V1179Her_corner_VR} and Fig. \ref{fig:V1179Her_corner_VI}. The averaged effective temperatures 
$T_{1} = 6171 \pm 135K$ and $T_{2} = 6034 \pm 117K$ for the primary and secondary respectively, are consistent with the binary temperature provided by LAMOST DR4.
The binary has a low mass-ratio $q = 0.153 \pm 0.001$, $\Omega = 2.067 \pm 0.04$, an inclination $i = 77\fdg2 \pm 0\fdg6$ and a moderate fill-out factor of 
$f=45\%$. The light curve and the residuals are shown in Fig. \ref{fig:lc_V1179Her}. The models colour curves are in good agreement with the observed ones.
The residuals are plotted as the flux ratio of the $V$ observations to the model flux calculated with the results of the $VI_{c}$ run.

\begin{table*}
\caption{Light curve solutions from the provisional analysis.\label{tabPhotSol}}
\begin{tabular}{lccc}
\hline
                            & TYC 3700-1384-1    &       V1511 Her      & V1179 Her          \\
\hline
$g_{1}=g_{2}$               &     0.32           &         0.32         &        0.32        \\       
$A_{1}=A_{2}$               &     0.5            &         0.5          &        0.5         \\
$x_{1bol}$ = $x_{2bol}$     &     0.657          &         0.650        &        0.649       \\
$y_{1bol}$ = $y_{2bol}$     &     0.189          &         0.139        &        0.156       \\
$x_{1V}$ = $x_{2V}$         &     0.698          &         0.729        &        0.728       \\
$y_{1V}$ = $y_{2V}$         &     0.287          &         0.274        &        0.273       \\
$x_{1R}$ = $x_{2R}$         &     0.605          &         0.636        &        0.636       \\
$y_{1R}$ = $y_{2R}$         &     0.287          &         0.278        &        0.278       \\
$x_{1I}$ = $x_{2I}$         &     0.514          &         0.545        &        0.544       \\
$y_{1I}$ = $y_{2I}$         &     0.270          &         0.263        &        0.263       \\ 
$T_1$                       &     6750 K         &         6096 K       &        6249 K      \\
                            &                    &                      &                    \\ 
$T_2$                       &     6576 K         &         6107 K       &        6085 K      \\
$q$ = $m_{2}/m_{1}$         &     0.192          &         0.132        &        0.143       \\
$i$                         &     82.07          &         74.66        &        77.05       \\
$\Omega$                    &     2.152          &         1.984        &        2.023       \\
$\Omega_i$                  &     2.214          &         2.054        &        2.084       \\
$\Omega_o$                  &     2.091          &         1.968        &        1.991       \\
$f$                         &     50\%           &         80\%         &        66\%        \\
                            &                    &                      &                    \\
$\chi^2_{r}$                &     3.01           &         2.34         &        6.18        \\

\hline
\end{tabular}
\end{table*}

\begin{table*}
\caption{Best fit models from the MCMC sampling. \label{tab:mmcposteriors}}
\begin{tabular}{ccccccccc}
\hline
Run           &  $T_{1}$                          &  $T_{2}$                       &  $q$                                     &  $i$                                 &  $\Omega$                           &  $\Omega_{i}$  &  $\Omega_{o}$  &  $f$     \\
\hline
\multicolumn{9}{|c|}{TYC 3700-1384-1} \\			  
\hline			  
$VR_{c}$      &  $6529\substack{+3 \\ -2} K$      &  $6400\substack{+4 \\ -6} K$   &  $0.1823 \substack{+0.0008 \\ -0.0012}$  &  $81\fdg4\substack{+0.3 \\ -0.2}$    &  $2.133\substack{+0.003 \\ -0.005}$ &  $2.188$       &  $2.071$       &  $47\%$  \\
$VI_{c}$      &  $6662\substack{+2 \\ -2} K$      &  $6545\substack{+5 \\ -7} K$   &  $0.1816\substack{+0.001 \\ -0.0007}$    &  $80\fdg4\substack{+0.2 \\ -0.2}$    &  $2.127\substack{+0.004 \\ -0.003}$ &  $2.186$       &  $2.070$       &  $50\%$  \\
average 	  &  $6596\pm 98 K$                   &  $6472\pm 106K$                &  $0.182\pm 0.001$                        &  $80\fdg9\pm 0.7$                    &  $2.130\pm 0.004$                   &  $2.187$       &  $2.071$       &  $49\%$  \\
\hline
\multicolumn{9}{|c|}{V1511 Her} \\
\hline			  
$VR_{c}$      &  $6730\substack{+10 \\ -10} K$    &  $6650\substack{+20 \\ -20} K$ &  $0.155\substack{+0.002 \\ -0.002}$      &  $76\fdg8\substack{+0.4 \\ -0.4}$    &  $2.067\substack{+0.003 \\ -0.005}$ &  $2.117$       &  $2.017$       &  $49\%$  \\
$VI_{c}$      &  $6856\substack{+6 \\ -6} K$      &  $6780\substack{+10 \\ -20} K$ &  $0.154\substack{+0.002 \\ -0.002}$      &  $76\fdg9\substack{+0.4 \\ -0.5}$    &  $2.065\substack{+0.008 \\ -0.009}$ &  $2.114$       &  $2.015$       &  $49\%$  \\
average 	  &  $6793\pm 101 K$                  &  $6715\pm 103K$                &  $0.154\pm 0.001$                        &  $76\fdg8\pm 0.1$                    &  $2.066\pm 0.001$                   &  $2.114$       &  $2.014$       &  $48\%$  \\
\hline
\multicolumn{9}{|c|}{V1179 Her} \\
\hline			  
$VR_{c}$      &  $6080\substack{+4 \\ -5} K$      &  $5956\substack{+4 \\ -8} K$   &  $0.152\substack{+0.002 \\ -0.002}$      &  $77\fdg6 \substack{+0.2 \\ -0.1}$   &  $2.064\substack{+0.004 \\ -0.007}$ &  $2.108$       &  $2.010$       &  $45\%$  \\
$VI_{c}$      &  $6263\substack{+3 \\ -3} K$      &  $6113\substack{+8 \\ -8} K$   &  $0.154\substack{+0.002 \\ -0.002}$      &  $76\fdg7\substack{+0.2 \\ -0.2}$    &  $2.070\substack{+0.007 \\ -0.004}$ &  $2.113$       &  $2.014$       &  $44\%$  \\
average 	  &  $6171\pm 135 K$                  &  $6034\pm 117K$                &  $0.153\pm 0.001$                        &  $77\fdg2\pm 0.6$                    &  $2.067\pm 0.004$                   &  $2.111$       &  $2.012$       &  $45\%$  \\
\hline
\end{tabular}
\end{table*}

\section{Eclipse Timings and Orbital Period Study}
Times of minimum light are calculated for the observations presented in this paper, and for publicly available observations from Super Wide Angle Search for Planets \citep[SWASP;][]{butters10}, 
Northern Sky Variability Survey \citep[NSVS;][]{wozniak04}, All-Sky Automated Survey All Star Catalogue \citep[ASAS-3;][]{pojmanski02} and 
All Sky Automated Survey for SuperNovae \citep[ASAS-SN;][]{shappee14}, if available for the studied star.
For V1179 Her also times of minima from \citet{diethelm07, diethelm08,diethelm10} and \citet{hubscher16} are included.
\\Many studies use the Kwee-van Woerden method \citep[KVW;][]{kwee56} to calculate the times of minimum light. The KVW method
assumes symmetric minima which is not always the case for binaries exhibiting star spot activity. Moreover, it is well known that this method
underestimates the uncertainties. In order to get more accurate uncertainty estimates, the times of minimum light are calculated with following
procedure. For the observations presented in this paper, and the time series observations from SWASP, first the times of minimum light are estimated by eye
from the light curves. Next, the observations limited to a time interval of $\pm 0.25P$ around the estimated time of minimum light, are fitted
to the obtained model light curves with {\sc phoebe} by shifting them in time, and also in magnitude for the SWASP observations. By limiting the
observations to a time interval of $\pm 0.25P$ around the estimated minimum, the times for individual primary and secondary minima can be
calculated separately. The timings from the observations presented in this paper are weighted averages of the minima calculated from the $V$, $R_{c}$ and $I_{c}$ light curves independently.
NSVS, ASAS-3 and ASAS-SN only provide a few observations nightly. Therefore the observations for a complete observation season have been phased
before fitting them to the model light curves. \\
The uncertainty of the obtained time of minimum light is estimated by the weighted median difference in time between the observations and the model light curve. The weights are assigned
based on the local slope of the model curve, assigning small or even zero weight to the observations around minimum and maximum light and allow for a reliable uncertainty estimate on the time of minimum
light. In order to demonstrate this, the reduced $\chi^2_{r}$ values are calculated from a linear regression fit to the times of minima for V1511 Her, obtained for the three seasons
of SWASP data separately. The reduced $\chi^2_{r}$ statistic for the 76 times of minima in the timespan of $2453141 < HJD < 2453272$, the 16 times of minima
in the timespan $2454297 < HJD < 2454336$ and the 60 times of minima in the time period of $2454586 < HJD < 2454688$ are $\chi^2_{r} = 1.65$, $\chi^2_{r} = 2.17$ and $\chi^2_{r} = 0.74$
respectively.

\subsection{TYC 3700-1384-1}
\begin{figure} 
\includegraphics[width=\columnwidth, angle=0]{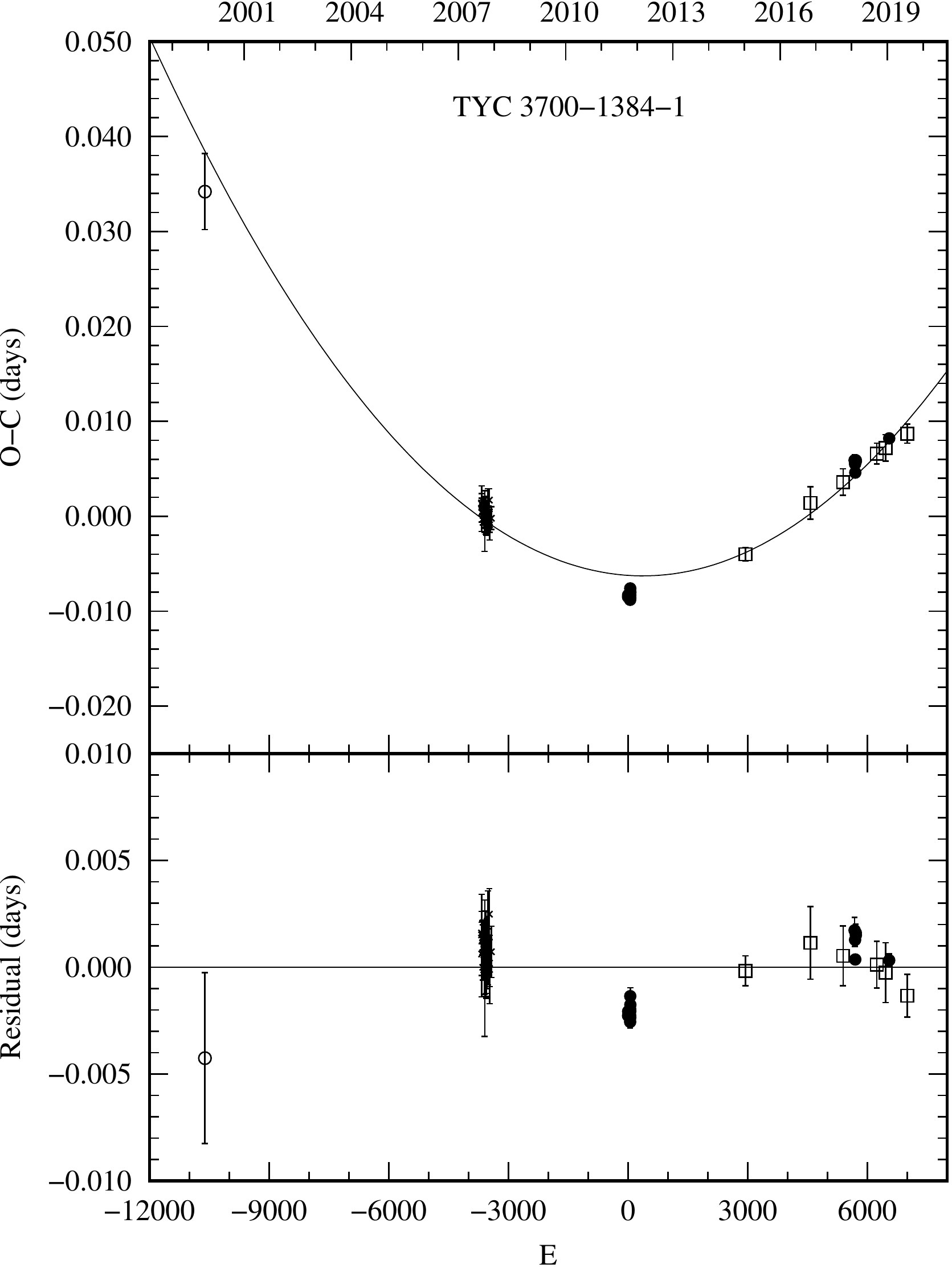}
\caption{
$O-C$ diagram of TYC 3700-1384-1. Open circles, NSVS data; crosses, SWASP data; open squares ASAS-SN data; filled circles, this paper.}
\label{fig:TYC_3700_oc}
\end{figure}

\citet{gettel06} derived a period of 0\fd40745944 from the NSVS observations but no detailed period study is published so far. Table \ref{tabToM_TYC3700-1384-1} lists the times of minimum light
calculated from NSVS, ASAS-SN, SWASP and the observations presented in this paper. From these timings the following least-squares ephemeris is obtained:
\begin{equation}
HJD~Min~I = 2455835.4930(9) + 0\fd4074723(2)\times{E}
\end{equation}
The $O-C$ residuals calculated with these ephemeris are plotted in Fig. \ref{fig:TYC_3700_oc}. The $O-C$ diagram indicates a linearly increasing period. A quadratic
least-square fit to the minima light times yields following ephemeris:
\begin{eqnarray}
HJD~Min~I = 2455835.4868(3) + 0\fd40747202(5)\times{E}\nonumber\\
+ 3.7(1)\times10^{-10}\times{E}^2
\end{eqnarray}
From the quadratic term, a continuous period increase rate of ${{\mathrm{d}}P/{\mathrm{d}}t}~=~6.1\times10^{-7}~d~yr^{-1}$, or $0.05~s~yr^{-1}$ is derived.

\subsection{V1511 Her}
\begin{figure}
\includegraphics[width=\columnwidth, angle=0]{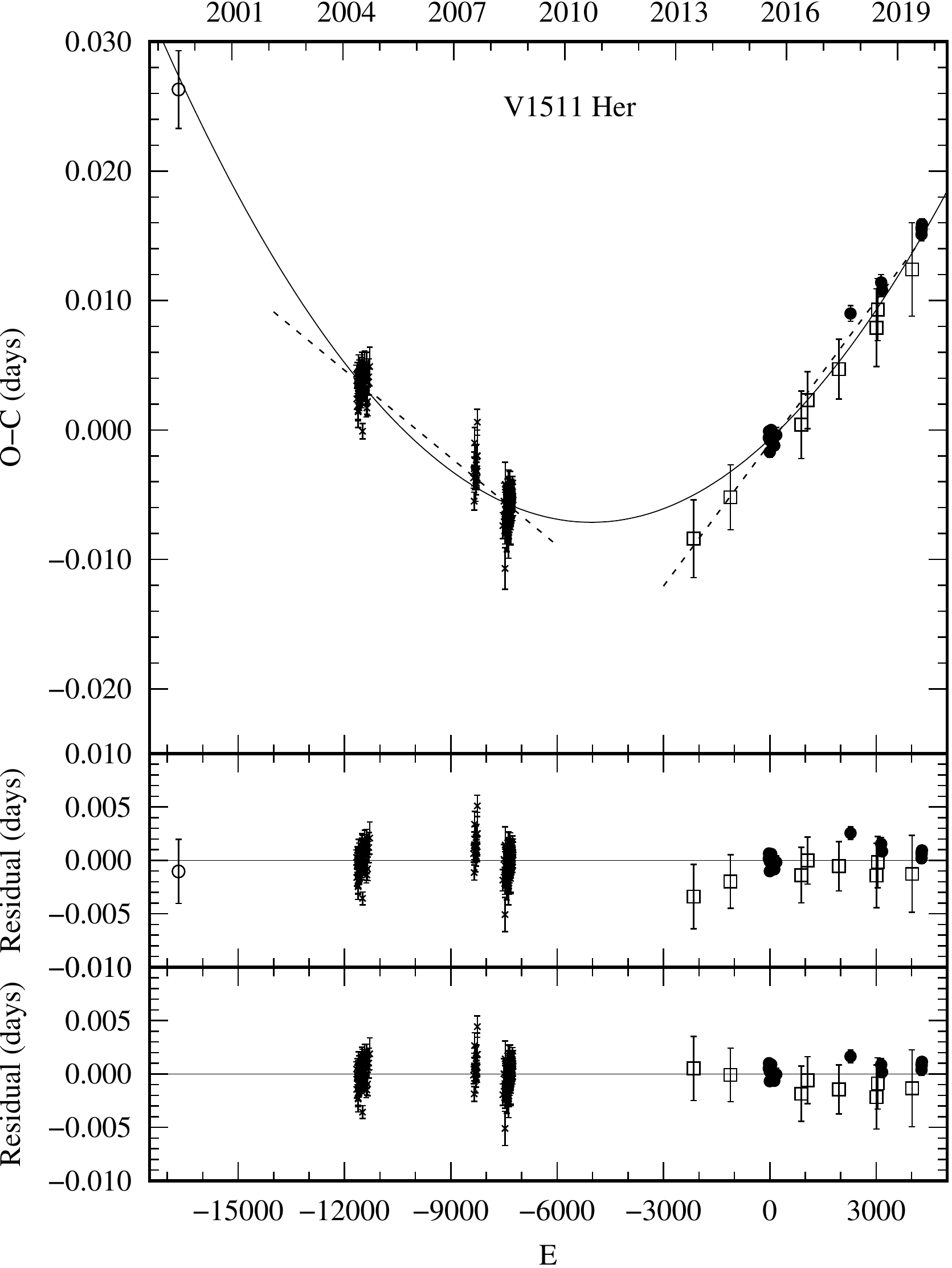}
\caption{
$O-C$ diagram of V1511 Her. The dashed lines refer to the linear ephemeris given in equations \ref{eq:V1511Her_L1} and \ref{eq:V1511Her_L2}.
Middle panel, residuals of the O-C values with respect to the quadratic ephemeris; lower panel, residuals with respect to the linear ephemerides. Symbols are the same as in Fig. \ref{fig:TYC_3700_oc}.}
\label{fig:V1511Her_oc}
\end{figure}

V1511 Her was discovered as a contact binary by \citet{akerlof00} and they published a period of $0\fd35005 \pm 0\fd00004$ based on ROTSE-I observations between
March 15th and June 15th 1999. \citet{gettel06} published a period of 0\fd3500768 based on the complete dataset of ROTSE-I observations spanning about 1 year. 
Observations of this star are publicly available from NSVS, ASAS-SN and SWASP. The times of minimum light obtained from these observations and
our observations are listed in Table \ref{tabToM_V1511Her}.
From all these timings we derived the following linear ephemeris:
\begin{equation}
HJD~Min~I = 2457223.4091(9) + 0\fd3500769(1)\times{E}
\end{equation}
The $O-C$ values based on this ephemeris plotted in Fig. \ref{fig:V1511Her_oc} suggests a continuously increasing period.
A quadratic fit to these times of minima yields following ephemeris:
\begin{eqnarray}
HJD~Min~I = 2457223.4084(2) + 0\fd35007946(5)\times{E}\nonumber\\
+ 2.54(4)\times10^{-10}\times{E}^2
\end{eqnarray}
implying a continuous period increase rate of ${{\mathrm{d}}P/{\mathrm{d}}t}~=~5.0\times10^{-7}~d~yr^{-1}$ or $0.04~s~yr^{-1}$.
No observations nor times of minima light are available for the period between 2000 and 2004, and between 2009 and 2013. We cannot exclude that
instead of a continuously increasing period, there are multiple sudden period changes. The times of minima between 2004 and 2009 and between 2013
and the present can be equally well fitted with respectively following linear ephemeris:
$-11660 \lid E \lid -7240$: \begin{equation} \label{eq:V1511Her_L1} HJD~Min~I = 2457223.3867(4) + 0\fd35007465(5)\times{E} \end{equation}
$-2150 \lid E \lid 4300$:   \begin{equation} \label{eq:V1511Her_L2} HJD~Min~I = 2457223.4081(3) + 0\fd3500806(1)\times{E} \end{equation}

\subsection{V1179 Her}
\begin{figure}
\includegraphics[width=\columnwidth, angle=0]{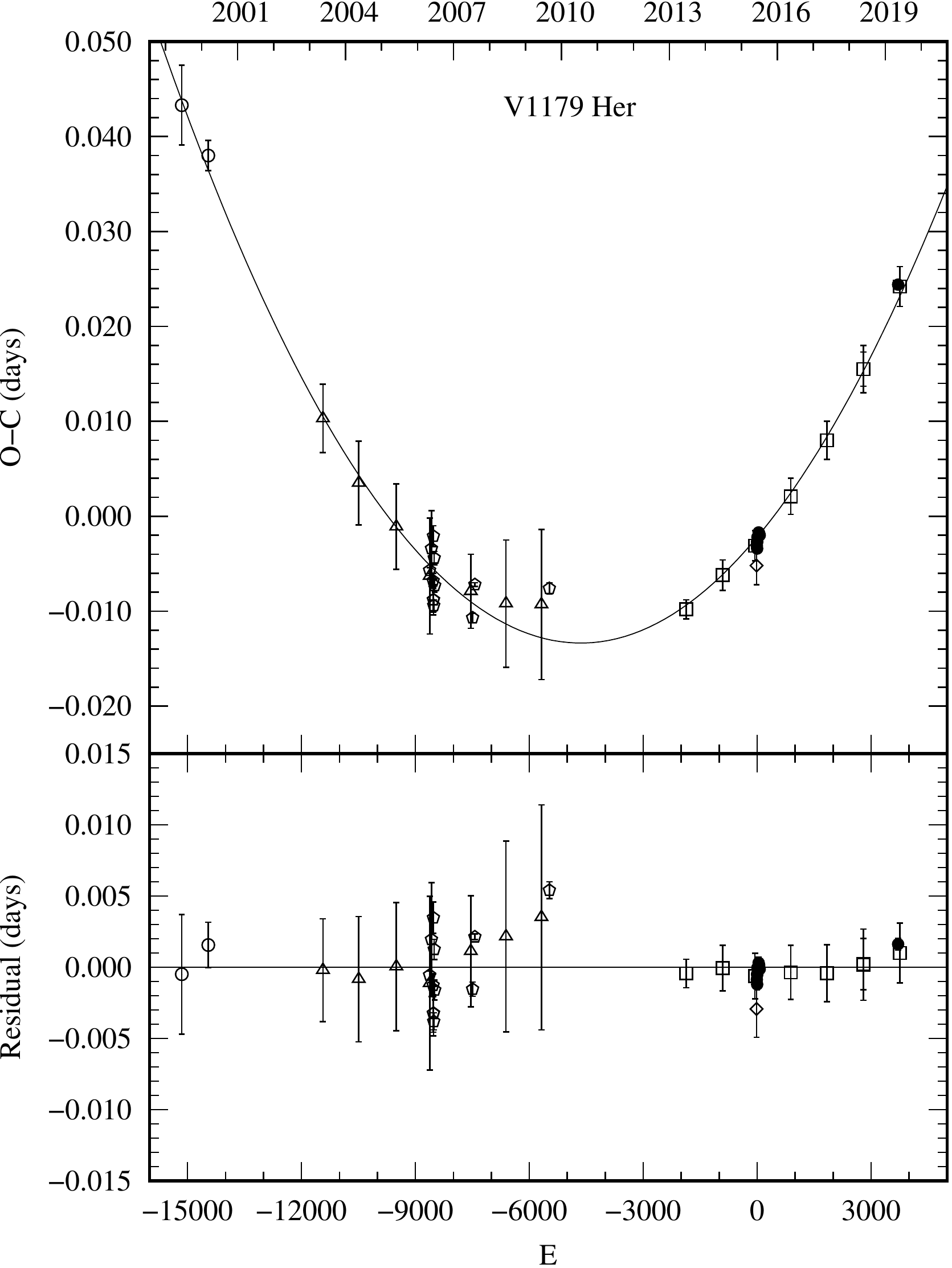}
\caption{
$O-C$ diagram of V1179 Her. Symbols are the same as in Fig. \ref{fig:TYC_3700_oc}, with additionally open triangles, ASAS-3 data; open pentagons, \citep{diethelm07,diethelm08,diethelm10};
open diamonds, \citep{hubscher16}.}
\label{fig:V1179Her_oc}
\end{figure}

V1179 Her has also been discovered by \citet{akerlof00} and they published a period of $0\fd38545 \pm 0\fd00003$. Since this star is at a declination of +11°,
the star has also been observed by the ASAS-3 from the Las Campanas Observatory in Chile. \citet{pojmanski02} published a period
of 0\fd385493 based on the ASAS-3 observations. The observations collected by NSVS, ASAS-3, ASAS-SN are publicly available. Unlike the two other stars
studied in this paper, no observations are available from SWASP. 
The calculated times of minimum light are listed in Table \ref{tabToM_V1179Her} together with those published by \citet{diethelm07,diethelm08,diethelm10} and \citet{hubscher16}.
Following linear ephemeris are derived from these observations: 
\begin{equation}
HJD~Min~I = 2457184.408(3) + 0\fd3855008(4)\times{E}
\end{equation}
The $O-C$ diagram plotted in Fig. \ref{fig:V1179Her_oc} clearly shows the signature of a linearly increasing period.
The quadratic ephemeris is:
\begin{eqnarray}
HJD~Min~I = 2457184.4058(4) + 0\fd3855056(1)\times{E}\nonumber\\
+ 5.2(1)\times10^{-10}\times{E}^2
\end{eqnarray}
implying a continuous period increase of ${{\mathrm{d}}P/{\mathrm{d}}t}~=~9.6\times10^{-7}~d~yr^{-1}$, or $0.08~s~yr^{-1}$.

\begin{table*}
\caption{Times of minima of V1511 Her. 
         \label{tabToM_V1511Her}}
\footnotesize
\begin{center}
\begin{tabular}{ccc|ccc|ccc}
\hline
Epoch    & HJD                     & data$^{\ast}$ & Epoch  & HJD                  & data$^{\ast}$ & Epoch  & HJD     & data$^{\ast}$\\
\hline
-16682.0  & 2451383.4526  $\pm$ 0.0030 & (1) & -11434.5  & 2453220.4599  $\pm$ 0.0010 & (4) &  -7364.0  & 2454645.4358  $\pm$ 0.0010 & (4) \\
-11659.5  & 2453141.6918  $\pm$ 0.0007 & (4) & -11426.0  & 2453223.4339  $\pm$ 0.0009 & (4) &  -7363.5  & 2454645.6124  $\pm$ 0.0011 & (4) \\
-11654.0  & 2453143.6153  $\pm$ 0.0008 & (4) & -11423.0  & 2453224.4850  $\pm$ 0.0006 & (4) &  -7361.0  & 2454646.4851  $\pm$ 0.0010 & (4) \\
-11634.0  & 2453150.6179  $\pm$ 0.0009 & (4) & -11417.5  & 2453226.4091  $\pm$ 0.0011 & (4) &  -7358.0  & 2454647.5361  $\pm$ 0.0012 & (4) \\
-11631.0  & 2453151.6684  $\pm$ 0.0009 & (4) & -11414.5  & 2453227.4606  $\pm$ 0.0006 & (4) &  -7352.0  & 2454649.6382  $\pm$ 0.0012 & (4) \\
-11628.5  & 2453152.5443  $\pm$ 0.0008 & (4) & -11411.5  & 2453228.5111  $\pm$ 0.0008 & (4) &  -7349.5  & 2454650.5147  $\pm$ 0.0010 & (4) \\
-11625.5  & 2453153.5937  $\pm$ 0.0011 & (4) & -11406.0  & 2453230.4364  $\pm$ 0.0006 & (4) &  -7346.5  & 2454651.5633  $\pm$ 0.0010 & (4) \\
-11622.5  & 2453154.6447  $\pm$ 0.0006 & (4) & -11403.0  & 2453231.4861  $\pm$ 0.0008 & (4) &  -7344.0  & 2454652.4377  $\pm$ 0.0010 & (4) \\
-11617.0  & 2453156.5677  $\pm$ 0.0011 & (4) & -11386.0  & 2453237.4376  $\pm$ 0.0005 & (4) &  -7335.0  & 2454655.5881  $\pm$ 0.0010 & (4) \\
-11614.0  & 2453157.6174  $\pm$ 0.0012 & (4) & -11383.0  & 2453238.4858  $\pm$ 0.0011 & (4) &  -7332.5  & 2454656.4660  $\pm$ 0.0010 & (4) \\
\hline
\end{tabular}
\begin{flushleft}
$^{\ast}$data: (1) NSVS \citep{wozniak04}, (3) ASAS-SN \citep{shappee14}, (4) SWASP \citep{butters10}, (5) this paper.\\
For minima observed in more than one passband the weighted mean of the timings in those passbands is listed.\\
A machine-readable version of this table is online available in its entirety with this publication. A portion is shown here for guidance regarding
its form and content.
\end{flushleft}
\end{center}
\end{table*}

\begin{table*}
\caption{Times of minima of TYC 3700-1384-1. 
         \label{tabToM_TYC3700-1384-1}}
\footnotesize
\begin{center}
\begin{tabular}{ccc|ccc|ccc}
\hline
Epoch    & HJD                     & data$^{\ast}$ & Epoch  & HJD                  & data$^{\ast}$ & Epoch  & HJD     & data$^{\ast}$\\
\hline
-10610.0  & 2451512.2461  $\pm$ 0.0040 & (1) & -3551.0  & 2454388.5581  $\pm$ 0.0010 & (4) &   53.5  & 2455857.2846  $\pm$ 0.0004 & (5) \\
 -3676.0  & 2454337.6262  $\pm$ 0.0005 & (4) & -3548.5  & 2454389.5782  $\pm$ 0.0010 & (4) &   54.0  & 2455857.4882  $\pm$ 0.0003 & (5) \\
 -3673.5  & 2454338.6448  $\pm$ 0.0019 & (4) & -3541.0  & 2454392.6328  $\pm$ 0.0008 & (4) &   54.5  & 2455857.6926  $\pm$ 0.0004 & (5) \\
 -3671.0  & 2454339.6626  $\pm$ 0.0020 & (4) & -3538.5  & 2454393.6524  $\pm$ 0.0008 & (4) &   56.0  & 2455858.3028  $\pm$ 0.0003 & (5) \\
 -3651.5  & 2454347.6090  $\pm$ 0.0008 & (4) & -3516.5  & 2454402.6174  $\pm$ 0.0022 & (4) &   56.5  & 2455858.5072  $\pm$ 0.0004 & (5) \\
 -3649.0  & 2454348.6270  $\pm$ 0.0006 & (4) & -3509.5  & 2454405.4690  $\pm$ 0.0007 & (4) & 2942.0  & 2457034.2725  $\pm$ 0.0007 & (3) \\
 -3646.5  & 2454349.6463  $\pm$ 0.0007 & (4) & -3507.0  & 2454406.4873  $\pm$ 0.0008 & (4) & 4572.0  & 2457698.4578  $\pm$ 0.0017 & (3) \\
 -3644.0  & 2454350.6636  $\pm$ 0.0006 & (4) & -3504.5  & 2454407.5064  $\pm$ 0.0010 & (4) & 5393.0  & 2458032.9947  $\pm$ 0.0014 & (3) \\
 -3641.5  & 2454351.6837  $\pm$ 0.0008 & (4) & -3477.5  & 2454418.5098  $\pm$ 0.0012 & (4) & 5681.5  & 2458150.5528  $\pm$ 0.0006 & (5) \\
 -3624.5  & 2454358.6102  $\pm$ 0.0012 & (4) & -3475.0  & 2454419.5262  $\pm$ 0.0011 & (4) & 5693.5  & 2458155.4420  $\pm$ 0.0003 & (5) \\
 -3619.5  & 2454360.6477  $\pm$ 0.0009 & (4) & -3472.5  & 2454420.5445  $\pm$ 0.0016 & (4) & 5698.0  & 2458157.2748  $\pm$ 0.0002 & (5) \\
 -3617.0  & 2454361.6659  $\pm$ 0.0006 & (4) & -3428.5  & 2454438.4740  $\pm$ 0.0012 & (4) & 5710.5  & 2458162.3695  $\pm$ 0.0004 & (5) \\
 -3614.5  & 2454362.6849  $\pm$ 0.0009 & (4) &     0.0  & 2455835.4845  $\pm$ 0.0004 & (5) & 5713.0  & 2458163.3880  $\pm$ 0.0003 & (5) \\
 -3600.0  & 2454368.5924  $\pm$ 0.0014 & (4) &     2.5  & 2455836.5034  $\pm$ 0.0004 & (5) & 6238.0  & 2458377.3118  $\pm$ 0.0011 & (3) \\
 -3592.5  & 2454371.6483  $\pm$ 0.0032 & (4) &    36.5  & 2455850.3573  $\pm$ 0.0004 & (5) & 6457.0  & 2458466.5488  $\pm$ 0.0014 & (3) \\
 -3568.0  & 2454381.6308  $\pm$ 0.0009 & (4) &    37.0  & 2455850.5611  $\pm$ 0.0004 & (5) & 6550.0  & 2458504.4448  $\pm$ 0.0003 & (5) \\
 -3565.5  & 2454382.6519  $\pm$ 0.0007 & (4) &    39.0  & 2455851.3761  $\pm$ 0.0003 & (5) & 7004.0  & 2458689.4377  $\pm$ 0.0010 & (3) \\
 -3563.0  & 2454383.6683  $\pm$ 0.0011 & (4) &    51.5  & 2455856.4693  $\pm$ 0.0004 & (5) &         &                            &     \\
 -3553.5  & 2454387.5409  $\pm$ 0.0008 & (4) &    52.0  & 2455856.6728  $\pm$ 0.0003 & (5) &         &                            &     \\
\hline
\end{tabular}
\begin{flushleft}
$^{\ast}$data: (1) NSVS \citep{wozniak04}, (2) ASAS-3 \citep{pojmanski02}, (3) ASAS-SN \citep{shappee14}, (5) this paper.\\
For minima observed in more than one passband the weighted mean of the timings in those passbands is listed.
\end{flushleft}
\end{center}
\end{table*}

\begin{table*}
\caption{Times of minima of V1179 Her. 
         \label{tabToM_V1179Her}}
\footnotesize
\begin{center}
\begin{tabular}{ccc|ccc|ccc}
\hline
Epoch    & HJD                     & data$^{\ast}$ & Epoch  & HJD                  & data$^{\ast}$ & Epoch  & HJD     & data$^{\ast}$\\
\hline
-15152.0  & 2451343.3432  $\pm$ 0.0042 & (1) & -8503.0  & 2453906.4903  $\pm$ 0.0007 & (6) &   -2.5  & 2457183.4415  $\pm$ 0.0006 & (5) \\
-14453.0  & 2451612.8029  $\pm$ 0.0016 & (1) & -8492.5  & 2453910.5352  $\pm$ 0.0007 & (6) &    0.0  & 2457184.4046  $\pm$ 0.0004 & (5) \\
-11435.0  & 2452776.2167  $\pm$ 0.0036 & (2) & -7540.0  & 2454277.7241  $\pm$ 0.0039 & (2) &    3.0  & 2457185.5618  $\pm$ 0.0004 & (5) \\
-10492.0  & 2453139.7371  $\pm$ 0.0044 & (2) & -7494.0  & 2454295.4543  $\pm$ 0.0005 & (7) &    8.0  & 2457187.4898  $\pm$ 0.0003 & (5) \\
 -9503.0  & 2453520.9928  $\pm$ 0.0045 & (2) & -7434.5  & 2454318.3951  $\pm$ 0.0002 & (7) &   39.0  & 2457199.4408  $\pm$ 0.0004 & (5) \\
 -8627.5  & 2453858.4941  $\pm$ 0.0003 & (6) & -6613.0  & 2454635.0820  $\pm$ 0.0067 & (2) &   52.0  & 2457204.4522  $\pm$ 0.0004 & (5) \\
 -8622.0  & 2453860.6138  $\pm$ 0.0061 & (2) & -5680.0  & 2454994.7542  $\pm$ 0.0079 & (2) &   67.5  & 2457210.4273  $\pm$ 0.0004 & (5) \\
 -8578.5  & 2453877.3860  $\pm$ 0.0040 & (6) & -5473.5  & 2455074.3618  $\pm$ 0.0006 & (8) &  884.0  & 2457525.1928  $\pm$ 0.0019 & (3) \\
 -8534.0  & 2453894.5374  $\pm$ 0.0008 & (6) & -1866.0  & 2456465.0537  $\pm$ 0.0010 & (3) & 1842.0  & 2457894.5085  $\pm$ 0.0020 & (3) \\
 -8529.0  & 2453896.4629  $\pm$ 0.0013 & (6) &  -909.0  & 2456833.9816  $\pm$ 0.0016 & (3) & 2794.0  & 2458261.5127  $\pm$ 0.0018 & (3) \\
 -8524.0  & 2453898.3971  $\pm$ 0.0011 & (6) &   -59.0  & 2457161.6604  $\pm$ 0.0016 & (3) & 2799.0  & 2458263.4402  $\pm$ 0.0025 & (3) \\
 -8523.5  & 2453898.5831  $\pm$ 0.0016 & (6) &   -15.5  & 2457178.4275  $\pm$ 0.0020 & (9) & 3712.0  & 2458615.4114  $\pm$ 0.0004 & (5) \\
 -8518.5  & 2453900.5100  $\pm$ 0.0006 & (6) &    -7.5  & 2457181.5136  $\pm$ 0.0005 & (5) & 3761.0  & 2458634.3007  $\pm$ 0.0021 & (3) \\
\hline
\end{tabular}
\begin{flushleft}
$^{\ast}$data: (1) NSVS \citep{wozniak04}, (2) ASAS-3 \citep{pojmanski02}, (3) ASAS-SN \citep{shappee14}, (4) SWASP \citep{butters10}, (5) this paper, (6) \citep{diethelm07}, (7) \citep{diethelm08}, (8) \citep{diethelm10}, (9) \citep{hubscher16}.\\
For minima observed in more than one passband the weighted mean of the timings in those passbands is listed.
\end{flushleft}
\end{center}
\end{table*}

\section{Results and Conclusion}
\begin{figure}
\includegraphics[width=\columnwidth, angle=0]{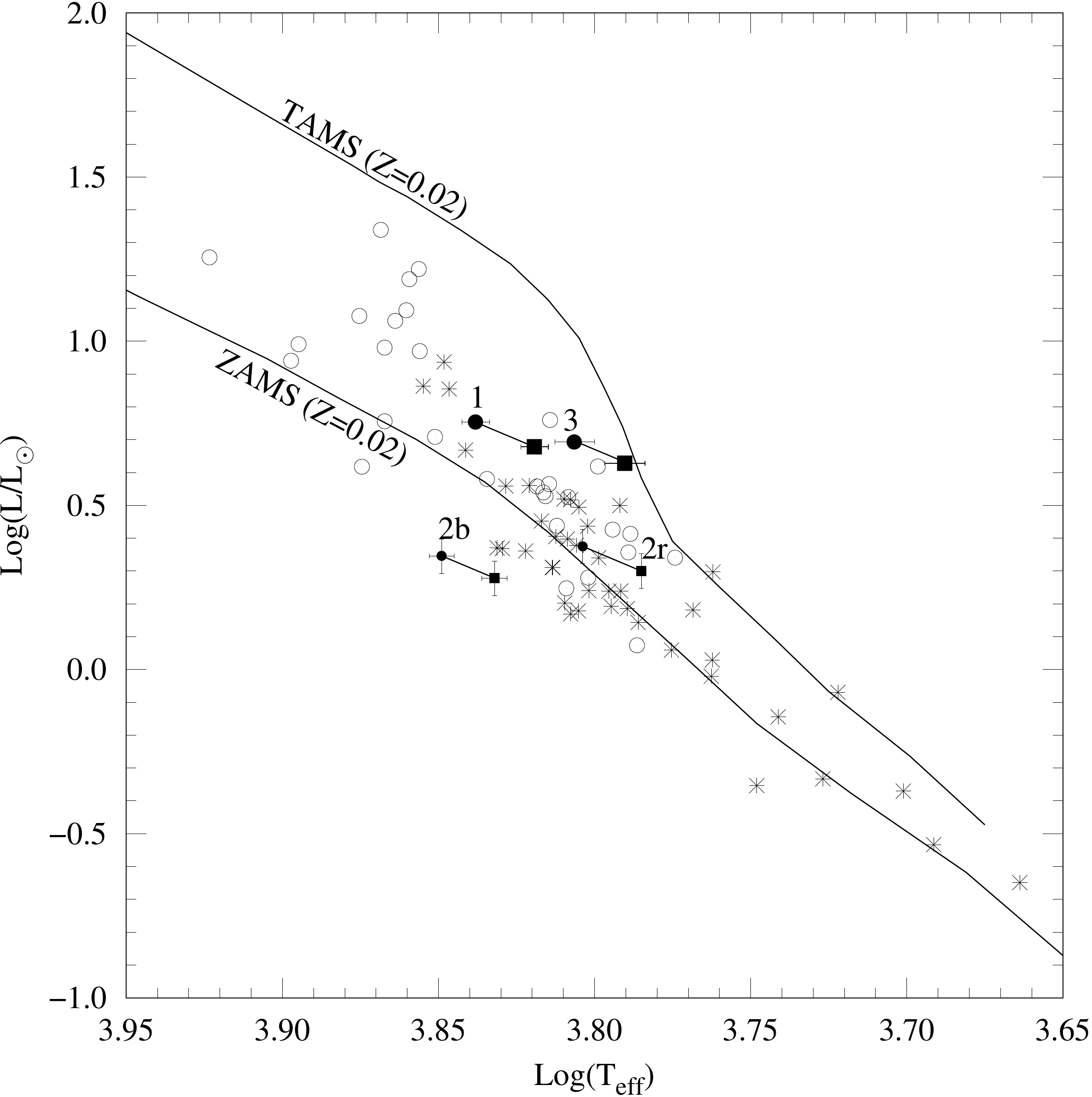}
\caption{
Positions of A-subtype (open circles) and W-subtype (asterisks) primaries, corrected for energy transfer, in the HR diagram. The positions of the stars discussed in this
paper (1, TYC 3700-1384-1; 2r \& 2b, V1511 Her; 3, V1179 Her) are plotted with filled circles and connected to filled squares which depict the positions without the correction for energy transfer.
The solid lines are the ZAMS and TAMS from \citet{girardi00}.}
\label{fig:HR}
\end{figure}

Photometric observations were obtained in the $V$, $R_c$ and $I_c$ passbands for the three W UMa stars TYC 3700-1384-1, V1511 Her and V1179 Her.
Since the orbital planes of the observed W UMa stars are aligned closely enough to our line-of-sight to manifest total eclipses, and the 
primary and secondary eclipse depths are similar indicating a small temperature difference between both components, stellar parameters can be determined
reliably from the photometric observations. Comparison star magnitudes from APASS DR9 were used. The Sloan $r'$ and $i'$ magnitudes given in APASS are transformed to the
Johnson-Cousins $V-R_{c}$ and $R_{c}-I_{c}$ colours using the equations from \citet{jester05}. The uncertainties, $\sigma_{r'} = 0.1$ and $\sigma_{i'} = 0.1$, for the comparison stars in
the fields of TYC 3700-1384-1 and $\sigma_{i'} = 0.15$ in the field of V1179 Her are large. Despite these large uncertainties, the colour indices for TYC 3700-1384-1 and
V1179 Her presented in this paper are in excellent agreement with the ones obtained by \citet{terrell12} using all-sky photometry. 
The uncertainties for the comparison stars in the field of V1511 Her, $\sigma_{V} \sol 0.02$, $\sigma_{i'} \approx 0.04$ and $\sigma_{r'} \approx 0.04$ are much smaller,
but the field has only been observed on 2 nights resulting in 10 measurements for each star. In contrast to these small uncertainties, the resulting $V-R_{c}$ and $V-I_{c}$ colour
indices for V1511 Her are inconsistent with those from \citet{terrell12}. 
\\
The light curves in the three passbands were initially modelled simultaneously using the WD method implemented in the legacy version of {\sc phoebe}. These resulting provisional 
system parameters were subsequently used in {\sc phoebe 2} for further analysis. The model passband fluxes were colour constrained
in order to derive temperatures for both components individually, without a priori assumptions, from the photometric observations transformed to the Johnson-Cousins standard system.
The photometric observations were converted to fluxes using a passband independent magnitude $m0$, and next in {\sc phoebe 2} scaled according to the \citet{castelli03} passband responses.
Although {\sc phoebe 2.2} supports interstellar reddening and extinction \citep{jones20}, the photometric measurements were de-reddened using table 3 in \citet{cardelli89}.
The uncertainties on the derived system parameters are estimated in {\sc phoebe 2} using the Markov Chain Monte Carlo (MCMC) method implemented via {emcee}. 
Since the observations have been made in 3 passbands, the temperatures are overdetermined. Therefore for each star two MCMC runs were executed, one with the $VR_{c}$ observations and one
with the $VI_{c}$ observations. The results are summarised in Table \ref{tab:mmcposteriors}.
The derived temperatures show that all three systems are of F-type spectral class. The temperature differences between the components for all three stars are small as one can expect
from the nearly equal primary and secondary eclipse depths. This makes the obtained system parameters less prone to the model discontinuities in the contact region of the binary \citep{kochoska18}. 
For all three systems the more massive component is eclipsed during the primary eclipse, hence the stars are of A-subtype according to the \citet{binnendijk70} classification.
\\
For TYC 3700-1381-1 the derived effective temperatures for the primary and secondary are $T_{1} = 6596 \pm 98K$ and $T_{2} = 6472 \pm 106K$ respectively. The temperatures 
and are in good agreement with the effective binary temperature ${\Teff} = 6500 \pm 125K$ estimated from the SED using a wide range of wavelengths. The uncertainties
on these temperatures are consistent with the uncertainties on the observed colour indices.
The small mass-ratio $q=0.182 \pm 0.001$ and dimensionless potential $\Omega = 2.130 \pm 0.004$ indicate a moderate fill-out factor of $f=49\%$. The derived orbital
inclination $i = 80\fdg9 \pm 0\fdg7$. The uncertainties are estimated from the standard deviation of the values provided by the two MCMC runs in {\sc phoebe 2}. The resulting model light
and colour curves from the best solution of both runs are plotted in Fig. \ref{fig:lc_TYC3700}. Because ${\Teff}$ is overdetermined with more than 2 bandpasses, one of the model's
colour curves shows a mismatch in flux level with respect to the observations. This is as expected, unless in the unlikely case that all systematic errors are
negligible small. As can be seen in Fig. \ref{fig:lc_TYC3700} the observed colour variations are well reproduced by the model.
\\
As shown in Fig. \ref{fig:lc_V1179Her}, the model light and colour curves for V1179 Her are also in good agreement with the observations. The averaged system parameters from the two MCMC runs are:
$T_{1} = 6171 \pm 135 K$, $T_{2} = 6034 \pm 117K$, $q = 0.153 \pm 0.001$, $\Omega = 2.067 \pm 0.04$ and $i = 77\fdg2 \pm 0\fdg6$. The fill-out factor $f = 45\%$. 
The temperatures are consistent with the ${\Teff} = 6249 \pm 12 K$ given in LAMOST DR4 for the binary.  The {\sc phoebe 2} MCMC posterior distributions for the run
with the $VR_{c}$ and $VI_{c}$ light curves are plotted in Fig. \ref{fig:V1179Her_corner_VR} and Fig. \ref{fig:V1179Her_corner_VI} respectively.
\\
While the uncertainties of the APASS comparison stars in the field of V1511 Her are reasonable, our measured colour indices are inconsistent with the ones provided by \citet{terrell12}.
The reddening free $(V-I_{c})_0 = 0.61$ from the observations of \citet{terrell12} suggest a temperature that is in good agreement with ${\Teff} = 6096 \pm 17K$ provided by LAMOST DR4. 
However, the reddening free $(B-V)_{0} = 0.45$ colour index from \citet{terrell12} suggest a much higher binary temperature of ${\Teff} \sim 6500 K$.
From the MCMC run in {\sc phoebe 2}, effective temperatures $T_{1} = 6793 \pm 101 K$ and $T_{2} = 6715 \pm 103K$ are derived, which are significantly higher than the binary temperature given by
LAMOST DR4. The uncertainties provided here are consistent with the errors on the derived colour indices, but don't account for a possibly larger systematic error in the values
of the comparison stars provided in APASS DR9.
The derived mass-ratio $q = 0.154 \pm 0.001$ and the value of $\Omega = 2.066 \pm 0.001$, resulting in a fill-out factor $f=48\%$, is somewhat higher than the mass-ratio $q=0.13$ from
the preliminary solution with the primary temperature fixed to the temperature given by LAMOST DR4. The inclination $i = 76\fdg8 \pm 0\fdg1$. 
The amplitude of the model colour curve based on the results from the MCMC runs, plotted in Fig. \ref{fig:lc_V1511Her}, does not match the observed colour changes.
Also the model fitted with the legacy version of {\sc phoebe}, with $T_{1}$ fixed to the value ${\Teff} = 6096 K$ provided by LAMOST DR4, does not reproduce the amplitude of the colour changes
completely. The amplitude for the $V-I{c}$ colour curve is only $5 mmag$ larger. From this initial analysis a temperature of $T_2 = 6107 K$ is obtained for secondary, and the very low
mass ratio $q=0.13$ and $\Omega = 1.984$ imply a very high fill-out factor of $f=80\%$. With the derived temperatures, and the model colour curves, being inconsistent,
the results for this system are inconclusive.
\\
The period study based on the times of minimum light presented in this paper, and calculated from the publicly available observations and those available in the literature
show that all three stars have a linearly increasing period. The period of TYC 3700-1384-1 is increasing by ${{\mathrm{d}}P/{\mathrm{d}}t}~=~6.1\times10^{-7}~d~yr^{-1}$ which equals 
5 seconds per century, the period of V1511 Her by ${{\mathrm{d}}P/{\mathrm{d}}t}~=~5.0\times10^{-7}~d~yr^{-1}$ or 4 seconds per century and the
period of V1179 Her by ${{\mathrm{d}}P/{\mathrm{d}}t}~=~9.6\times10^{-7}~d~yr^{-1}$ or 8 seconds per century.
Due to the lack of minimum timings between 2000 and 2004 and between 2009 and 2013 it cannot be excluded that the period change of V1511 Her is composed of multiple discrete period changes. 
Linearly increasing periods are usually attributed to mass transfer from the less massive component to the more massive one through the neck of the system.
\\
Using the distance provided by $Gaia$ DR2 and taking the reddening obtained from the Bayestar19 3D dust map into account, the luminosities for
the primary components in these systems are calculated. The bolometric corrections are taken from the \citet{pecaut13} online table, version 2019.3.22. The nearly equal effective
temperatures of both components in W UMa stars despite their very different masses, is often explained by energy transfer from the more massive to the less massive star.
\citet{mochnacki81} defined a quantity $U$ as the transfered component of the secondary’s luminosity expressed as the fraction
of the energy radiated from the surface of the primary. This quantity allows to correct the luminosity, and the temperature of the primary at constant radius, for
the energy transfer in order to compare the primaries with zero-age main sequence stars. Fig. \ref{fig:HR} depicts the positions of the primaries in the HR diagram, and their
positions when corrected for the energy transfer to other A and W-subtype primaries taken from \citet{yakut05}. The filled squares indicate the positions with the observed properties and the
filled circles the positions corrected for energy transfer. The primary component of TYC 3700-1384-1 fits well in the main sequence, while the primary of V1179 Her appears to be more evolved.
Despite the analysis of V1511 Her is inconclusive, the primary is plotted, with smaller symbols, in Fig. \ref{fig:HR}. The uncertainty
on the $Gaia$ distance for this star is also \textasciitilde 5 times larger than for the other two studied stars, while the distances are similar. The position
in the HR diagram based on the provisional solution with $T_{1}$ fixed to the temperature from LAMOST DR4 is indicated with $2r$ in the plot, while the bluer solution derived from the observed colour
indices is indicated with $2b$. The bluer solution falls well below the ZAMS, hence favoring the provisional solution based on the cooler temperature from LAMOST DR4.
The primary temperatures corrected for energy transfer, $T^{c}_{1} = 6888K$ and $T^{c}_{1} = 6405 K$ for TYC 3700-1384-1 and V1179 Her respectively, suggest a primary mass of $M_{1} = 1.45 M_{\sun}$
for TYC 3700-1384-1 and $M_{1} = 1.3 M_{\sun}$ for V1179 Her. The mass transfer rates from the less massive component to the more massive one are estimated using these masses and the derived period
change rates.  For TYC 3700-1384-1 the mass transfer rates amounts to $\dot M=1.6 \times 10^{-7} {\Msun} yr^{-1} $ and for V1179 Her to $\dot M= 1.9 \times 10^{-7} {\Msun} yr^{-1} $. 

\section*{Acknowledgements}
The author would like to thank the anonymous referee for giving valuable suggestions and comments that resulted
in a significant improvement of the manuscript. Special thanks are due to Dr. Kyle Conroy for the support provided on using {\sc phoebe 2} to model the studied stars.
This research made use of the International Variable Star Index (VSX) data base, operated at AAVSO, Cambridge, Massachusetts, USA.
Additionally this study made use of NASA’s Astrophysics Data System, and the SIMBAD and VizieR data bases operated at the CDS, Strasbourg, France.

\section*{Data Availability}
The observations presented in this paper are available upon request from the author.
Part of this paper makes use of publicly available data collected by SuperWide Angle Search for Planets (\url{https://wasp.cerit-sc.cz/form}),
Northern Sky Variability Survey (\url{https://skydot.lanl.gov/nsvs/nsvs.php}), All-Sky Automated Survey All Star Catalogue (\url{http://www.astrouw.edu.pl/asas/?page=asas3}) and 
All Sky Automated Survey for SuperNovae (\url{https://asas-sn.osu.edu/}).

\bibliographystyle{mnras}
\bibliography{bib}

\label{lastpage}
\end{document}